\documentclass[fleqn,usenatbib]{mnras}

\usepackage[T1]{fontenc}
\usepackage{ae,aecompl}

\usepackage{graphicx}	
\usepackage{amsmath}	
\usepackage{amssymb}	

\usepackage{mathrsfs}

\usepackage{amsxtra}
\usepackage{textcomp}
\usepackage{afterpage}
\usepackage{cases}
\usepackage{pdflscape}
\usepackage{rotating}
\usepackage{color}
\usepackage{multirow}
\usepackage{natbib}
\usepackage{times}
\usepackage{natbib}
\usepackage{amsmath}
\usepackage{amssymb}
\usepackage{textcomp}
\usepackage[normalem]{ulem}
\usepackage{multirow}
\usepackage{capt-of}
\usepackage{array}
\usepackage{tikz}
\usetikzlibrary{trees}
\usepackage[colorinlistoftodos]{todonotes}
\usepackage{cases}
\usepackage[export]{adjustbox}

\title[SF in A3376]{Passive spirals and shock influenced star formation in the merging cluster A3376}

\author[K. Kelkar et al.]{Kshitija Kelkar$^{1}$\thanks{E-mail: kshitija@rri.res.in (KK)},
K. S. Dwarakanath$^{1}$,
Bianca M. Poggianti$^{2}$,
Alessia Moretti$^{2}$,
\newauthor Rogerio Monteiro-Oliveira $^{3}$,
Rubens Machado$^{4}$,
Gastao Lima-Neto$^{3}$,
Jacopo Fritz$^{4}$,
\newauthor Benedetta Vulcani$^{2}$, 
Marco Gullieuszik$^{2}$,
Daniela Bettoni$^{2}$
\\
$^{1}$Raman Research Institute, Bangalore, India\\
$^{2}$INAF - Astronomical Observatory of Padova, 35122 Padova, Italy\\
$^{3}$Instituto de Astronomia, Geof\'{i}sica e Ci\^{e}ncias Atmosf\'{e}ricas, Universidade de S\~{a}o Paulo, R. do Mat\~{a}o 1226, 05508-090 S\~{a}o Paulo, Brazil\\
$^{4}$Departamento Acad\^emico de F\'isica, Universidade Tecnol\'ogica Federal do Paran\'a, Rua Sete de Setembro 3165, Curitiba, Brazil\\
$^{5}$Instituto de Radioastronom\'{i}a y Astrof\'{i}sica, IRyA, UNAM, Campus Morelia, A.P. 3-72, C.P. 58089, Mexico\\
}

\date{Accepted XXX. Received YYY; in original form ZZZ}

\pubyear{2015}

\begin{document}
\label{firstpage}
\pagerange{\pageref{firstpage}--\pageref{lastpage}}
\maketitle

\begin{abstract}
We present a detailed analysis of star formation properties of galaxies in a nearby ($z\sim0.046$) young ($\sim$0.6 Gyr) post-merger cluster system A3376, with a moderate shock front ( $v_{s}\sim$1630 km/s) observed as symmetric radio relics. Exploiting the spectroscopic data from the wide-field OmegaWINGS survey and the associated photometric information, our investigations reveal the plausible effects of the dynamic post-merger environment differing from the putative pre-merger cluster environment. The remnants of the pre-merger relaxed cluster environment are realised through the existence of passive spiral galaxies located in the central regions of the cluster between the two BCGs. We discover A3376 to contain a population of massive $(M_{*}/M_{\odot})>10$) blue regular star-forming spirals in regions of maximum merger shock influence but exhibiting star formation rates similar to those in relaxed clusters at similar epoch. We further discover low-mass (Log $(M_{*}/M_{\odot})\leq 10$) late-type blue PSBs which could either be formed as a result of rapid quenching of low-mass spirals following the shock-induced star formation or due to the intense surge in the ICM pressures at the beginning of the merger. With the possibility of the merger shock affecting high- and low-mass spirals differently, our results bridge the seemingly contradictory results observed in known merging cluster systems so far and establish that different environmental effects are at play right from pre- to post-merger stage.
\end{abstract}




\section{Introduction}
\label{intro}

Large scale structures like galaxy clusters undergo hierarchical growth with subsequent accretion of groups and cluster mergers throughout cosmic time. Recent developments in multiwavelength studies of galaxy clusters from the present epoch till intermediate redshifts have shown that high-density environments influence the star formation, gas content and structure of their galaxies. Cluster merging events, however, disrupt the cluster as a whole and their effects are evident on the intracluster medium, morphology of the system and the distribution of inherent galaxies. Merging galaxy clusters can hence lend a new insight into the influence of such a dynamic environment on their galaxy populations, possibly affecting the environmental trends observed in galaxy clusters today.

Dense cluster environment is found to have little effect on the main structural properties of the galaxies such as their sizes for a given mass and morphology, and their internal structure,  but it affects the morphological mix and star formation history of galaxies \citep[][and references within]{kelkar15,kk17}. Moreover it is widely observed that the cluster environment is harsh to the star formation, with star-forming cluster galaxies having reduced star formation rates and quiescent galaxies being overabundant in denser regions \citep{dressler1980, poggianti06,quadri12,fasano15}. Furthermore, the established morphological mix of cluster galaxies points towards a possible morphological transformation which galaxies undergo, albeit at timescales longer than those required to shut down star formation in them \citep{kelkar19}. The current consensus hence strongly favours gas removal mechanisms like ram-pressure stripping \citep{gunngott72}, and starvation \citep{larson80} to be the key processes in shutting down star formation in  cluster galaxies without bringing a global structural change in galaxies. Indeed mounting evidence for ram-pressure stripping is now revealed through observations of ongoing/recent stripping of cold gas in galaxies, coined popularly as `jellyfish' galaxies \citep{fumagalli14,fossati16, gasp2,gasp1,gasp9} as well as hydrodynamical N-body simulations such as \citet{ruggiero17}. 

The growth of galaxy clusters takes place through large-scale merging events, as predicted by the hierarchical paradigm for structure formation. Such Mpc-scale events release copious amounts of kinetic energy part of which is dissipated through non-thermal processes in the intracluster medium (ICM) often observed as diffuse synchrotron radio emission in the form of halos (located centrally in the merging cluster) and relics (detected towards cluster peripheries), and heating the ICM to give rise to thermal X-ray emission from the cluster. One likely mechanism giving rise to halos is believed to be the reacceleration of electrons to ultra-relativistic energies due to turbulence injected into ICM during mergers \citep{brunetti01} while relics are created as a result of the repeated diffusive acceleration of thermal particles across the merger shock front emanating from the cluster centre at onset of mergers \citep{feretti12}. While the recent merger history and dynamical state of clusters is revealed by the presence of diffuse radio emission and the morphology of X-ray emission, the detectability of such non-thermal indicators depends largely on the merger geometry, the mass ratio of the participating clusters, age of the merger, turbulence decay timescale, and synchrotron life-time of decaying electrons \citep{brunetti14}. Better low-frequency radio studies however are contributing to the growing number of cluster merger candidates through detections of radio relics and halos, and X-ray shock fronts observed at various stages of cluster merging \citep[eg.][]{akamatsu13,bourdin13,sarazin16, akamatsu17-b,gu19}. Other methods of discovering cluster merger candidates include optical analysis \citep[eg.][]{kaya19}, X-ray surface brightness techniques \citep[eg.][]{lyskova19}, IFU spectroscopy observations \citep[eg.][]{jauzac19}, and machine-learning techniques applied to galaxy catalogues \citep[eg.][]{delosrios16}.   

However the role of these merging events in setting the environmental trends, already observed in several galaxy properties in dynamically relaxed cluster environments, is still not very well understood. It is unclear whether such cluster mergers quench the star formation in galaxies \citep{mansheim17-b} or trigger it \citep{mansheim17-a}, as very few merger systems are studied till date with evidence supporting both scenarios. Contrary to the established quenching of star-forming activity occurring in cluster environments, several recent studies discover an unusual population of H$\alpha$ emitters in merging clusters, where the merger shock is believed to be responsible for substantial star formation for at least $\sim 100 Myr$ \citep{umeda04,stroe15a,stroe15b}. However, merger systems like the `Toothbrush' cluster reveal a possibility that perhaps the age of the merger holds the key to whether the shock-induced star formation, which gets rapidly quenched, would be detected or not \citep{stroe15-c}. Furthermore, growing evidence is presented for enhanced ram-pressure stripping initiated due to dense ICM in merging environment, by the incidence of `Jellyfish' galaxy candidates in the proximity of the shock front \citep{owers12,rawle14}, also demonstrated by recent hydrodynamical simulations by \citet{roediger14} and \citet{ruggiero19}. \citet{mansheim17-a} however, observe a suppressed star formation in merging cluster environments at $z\sim1$ which they argue to be caused by the gravitational tidal forces arising from merging halos. This observation is in line with recent simulation studies like \citet{bekki10} who report a significant incidence of galaxies in transition, like the post-starburst galaxies, being formed as a result of the sharp increase in ICM pressure during the merging event. In summary, this suggests that such energetic environments are far too complex to comprehend and the available systems far too less to connect observations to the dynamic properties of merging clusters.  

In order to address these paradigms, we choose a unique nearby merger system, Abell 3376, whose merger dynamics and ICM properties are well established through the wealth of multiwavelength data. This paper, however, will present the first ever in-depth analysis of the galaxy populations of Abell 3376 using the data from one of the widest spectroscopic cluster surveys, exploring implications of shocked ICM on the star formation properties of cluster galaxies and connecting them to the purported trends observed in known merging/post-merger systems. The paper is organised as follows: With a brief introduction to Abell 3376 system in Section \ref{intro}, Section \ref{data} describes the overall multiwavelength data available with Section \ref{s-galaxysample} introducing the galaxy sample used throughout this work. We present our results in Section \ref{results} where we start with exploring the general galaxy populations encountered in Abell 3376, and then focussing on star formation in this system through a subsample of star-forming spiral galaxies and post-starburst galaxies. We discuss the implication of our results in the global framework of galaxy trends observed in merging cluster systems, followed by Section \ref{conclusions} summarising our key results and conclusions. Throughout this paper, we use the standard $\Lambda$CDM cosmology ($h_0$=0.7, $\Omega_\Lambda$=0.7 and  $\Omega_m$=0.3), and \citet{salpeter55} initial mass function in the mass range 0.15-120 $M_\odot$.
 
\subsection{The merger system of A3376}

Abell 3376 (hereafter A3376) is a merging cluster at $z_{cl}=0.046$ \citep{struble99} discovered through disturbances observed in the ICM \citep{ebeling96,flin06}. This was further verified by the existence of two symmetric radio relics indicating the merger shock front $\sim 2$ Mpc apart \citep{bagchi06,kale12,lijo15} and a distinct cometary X-ray morphology \citep{akamatsu12}. Optically, projected galaxy overdensities were confirmed about the two Brightest cluster galaxies (BCGs) in A3376:  the BCG of the western (W) A3376 or `BCG W' (ESO307-13; RA=6$^h$00$^m$41$^s$.10, Dec=-40$^\circ$02$^\prime$40$^{\prime \prime}$.00) and the BCG of the eastern (E) A3376 or `BCG E' (2MASXJ06020973-3956597; RA=6$^h$02$^m$09$^s$.70, Dec=-39$^\circ$57$^\prime$05$^{\prime \prime}$.00) by \citet{ramella07} in the WINGS \citep{fasano06} data. In addition, \citet{mo17} also report a new galaxy concentration detected in their wide-field $R-$band data dedicated for weak-lensing studies, and is found to lie north of BCG E and the eastern relic (referred to as A3376 N with BCG N). A3376 also appears to display disturbed $B-$band luminosity function, expected from disturbed or merger cluster systems \citep{durret13}. 

N-body SPH simulations performed by \citet{machado13} have proposed that this merger is mostly occurring in the plane of the sky with a very small impact parameter ($\sim$few kpcs) making it a headlong merging system. They further propose that the compact dense cluster passed through the massive but sparser cluster, disrupting the ICM in its core $~0.5-0.6$ Gyrs ago. This scenario was corroborated by \citet{mo17} through weak lensing studies which show that the most significant mass peak is around the BCG W, away from the X-ray hotspot, with cluster masses measured to be M$^{W}_{200}\sim3.0\times10^{14}M_{\odot}$ and M$^{E}_{200}\sim0.9\times10^{14}M_{\odot}$. This makes it a 3:1 mass ratio cluster merger system observed at the present as the clusters are going to the point of farthest separation. Radio spectral index studies of the two relics estimate a Mach number ($M)\sim 2-3$ for the merger shock \citep{kale12,lijo15} in corroboration with past X-ray studies \citep{akamatsu12}. However, recent X-ray observations have revealed that the eastern shock is weaker ($M\sim1.5$) with shock speed $v_s$=1450 km/s than the western shock ($M\sim2.8$; $v_s$=1630 km/s), placing the dynamical age of the shock to $\sim0.6$ Gyr \citep{kale12,lijo15,urdampilleta18}.     

\section{Data}
\label{data}
Our analysis is based on the OmegaWINGS survey data for A3376 \citep{gull15,moretti17} which is one of the unique spectroscopic surveys having a 
spatial coverage of $\sim1$ sq degree, enabling us to investigate the wider environment of our merging system. Designed as an extension to the original WIde-field Nearby Galaxy-cluster Survey \citep[WINGS;][]{fasano06,moretti14} comprising 76 clusters, OmegaWINGS observed 57 of the WINGS clusters, the selection details of which are discussed in \citet{gull15}. These clusters have photometric and imaging data in the $U-$,$B-$,$V-$ bands using the OmegaCAM/VST, and a spectroscopic follow up for a subsample of 46 out of 57 clusters using AAOmega spectrograph at AAT \citep{moretti17}.   

The mean cluster redshift $z_{cl}$($\sim$0.0463), and the cluster velocity dispersion $\sigma_{cl}$($\sim$844 km/s) were iteratively determined through $3\sigma$ clipping using the biweight robust location and scale estimators (Beers 1990). The cluster membership was assigned to galaxies if they lie within $3\sigma_{cl}$ from the cluster redshift \citep{moretti17}. 

\begin{figure*}
	\includegraphics[width=\textwidth]{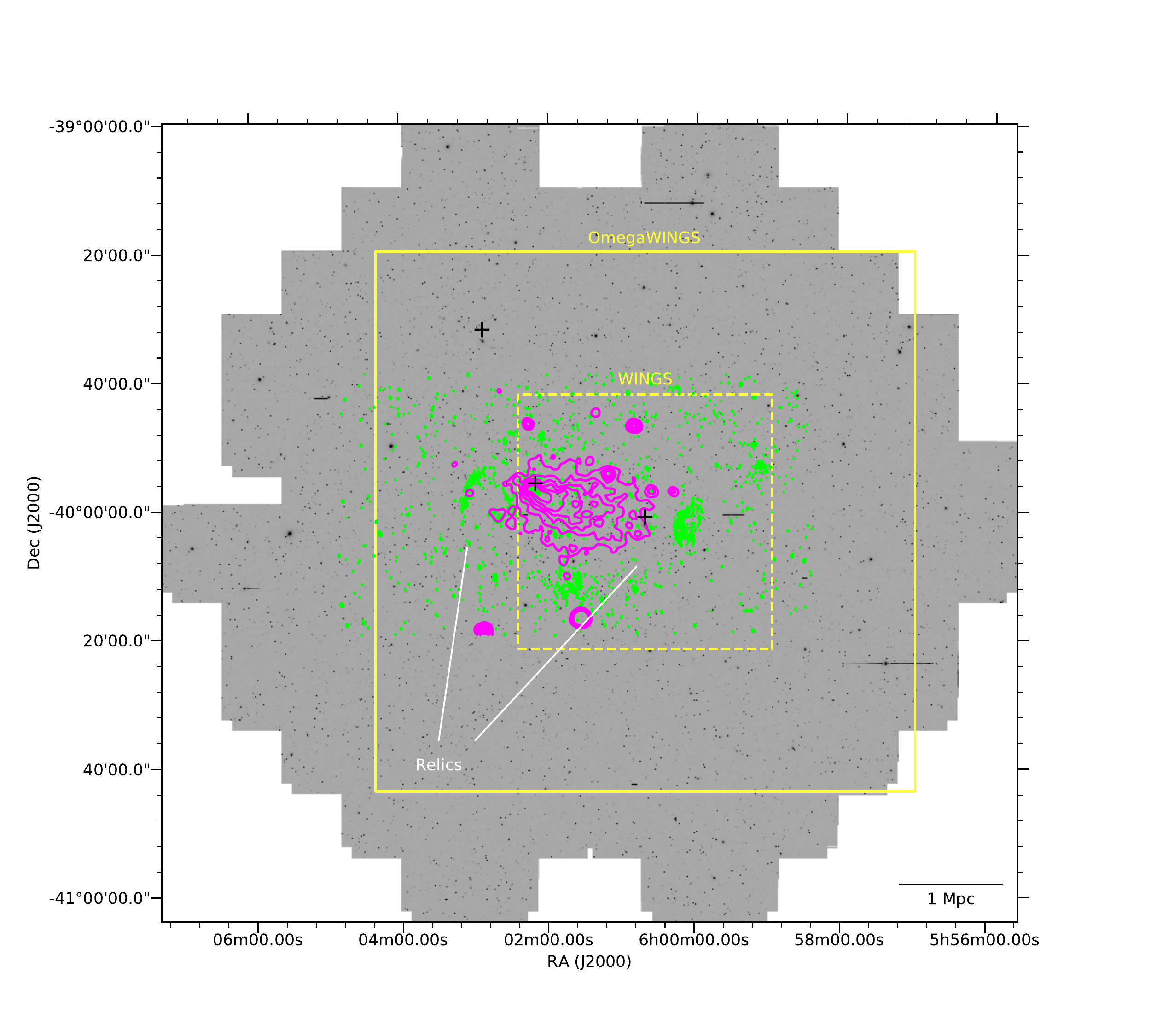}
    \caption{$R-$band mosaic of A3376, adapted from \citet{mo17}, overlaid with diffuse radio emission at 325MHz \citep[green,][]{kale12} and X-ray emission \citep[magenta,][]{akamatsu12}. The plus signs mark the positions of BCGs E and W along with BCG of the A3376 N group. The dashed and solid yellow squares denote the coverage of WINGS, and OmegaWINGS observations of the field, respectively. The arc-like like green contours near BCG E and BCG W are the radio relics showing the current position of the post-merger shock front.}
    \label{a3376_r}
\end{figure*}

Figure~\ref{a3376_r}  shows the $R-$band mosaic of A3376 obtained from Dark Energy Camera (DECAM) on the 4m Victor Blanco Telescope \citep{mo17} along with the current position of the shock front \citep[radio relics,][]{kale12}, and the Suzaku X-ray emission \citep{akamatsu12} relative to the positions of the BCGs. Overlaid on this image are the footprints of OmegaWINGS survey for A3376 highlighting the large spatial coverage in comparison with that of WINGS. 
\subsection{Morphology of galaxies}
\label{ss-morph}
Cluster galaxies from A3376 were morphologically categorized using the revised Hubble T-type morphological classification 
(\cite{fasano12} for WINGS, (Gianni Fasano private communication for OMEGAWINGS). These classifications were performed using a non-parametric automated tool called \textsc{morphot}, designed specifically for large galaxy surveys \citep{fasano07}, on the $V-$band imaging data. Along with the non-parametric quantitative structural classification (CAS), \textsc{morphot} assigns a morphological type (\textsc{morphot} type, $T_{M}$) to the galaxies ranging from -6 (cD) to 11 (Irregulars). We use the \textsc{morphot} WINGS+OMEGAWINGS morphologies for the galaxies in A3376 field, regrouped in three broad bins of Ellipticals (E:$-6< T_{M}\leq -4.25$), Lenticulars (S0:$-4.25 < T_{M}\leq 0$) and Spirals (S:$0<T_{M}\leq 8$).    

\subsection{Spectral classification and SFH of galaxies}
\label{sp-class}

The primary focus of this study is investigating the current star formation rates of galaxies in A3376, and characterise 
the plausible effects of large-scale cluster mergers and the resultant outgoing shocks on their star formation properties. 
For this purpose, we utilise the data byproducts given by SINOPSIS \citep{fritz07}, which is a spectrophotometric modelling code that derives several properties of stellar populations whose light we observe in the integrated spectrum of galaxies, namely the total stellar masses, star formation histories, magnitudes, extinction and equivalent widths of prominent spectral lines. The observed spectra refer to the light integrated within each fiber, and therefore they cover only the central few kpc of each galaxy. We applied appropriate correction for this effect, taking into account the measured color gradient \citep{fritz11}. Wherever applicable, the current star formation rate (SFR), i.e. within the last 0-20 Myr,  used throughout this paper is computed by fitting the equivalent widths of emission lines like $H\alpha$, $H\beta$ and [O\textsc{ii}] in galaxy spectra \citep{fritz11}. We assume a \citet{salpeter55} IMF, with masses in the range 0.15 to 120 M$_\odot$, and the total stellar mass was given using mass definition number 2 \citep[See][]{longhetti09}, that includes stars in the nuclear-burning phase and remnants, but takes into account mass losses due to stellar evolution.

We further apply the spectral classification scheme, first introduced by \citet{dressler99,poggianti99}, and later updated by \citet{fritz14}, to bifurcate galaxies based on the nature of their spectra and the presence of key line indicators. This classification scheme relies primarily on the rest-frame EWs of the [O\textsc{ii}] and H$\delta$, which are both good indicators of current and recent star formation (in past Gyr), respectively. Following the revision introduced by \citet{pacc17}, we first consider galaxies with H$\alpha$ in emission as  emission-line (`EML') galaxies. Galaxies with no H$\alpha$, but other very strong emission lines are also classified as EML galaxies. These galaxies are further classified into classical star-bursting systems (galaxies with e(b)-type spectra) or dusty starbursts/galaxies with abruptly truncated substantial star formation (e(a)) based on the strength of H$\delta$ in absorption. Regular star-forming galaxies (e(c)) are identified based on the presence of moderate-to-weak emission lines and moderate-to-weak H$\delta$ in absorption. Galaxies exhibiting a k-type spectra with no emission lines and weak H$\delta$ in absorption are  classified as passive/k-type galaxies which have no current or recent star formation. Galaxies showing no emission lines in their spectra but having strong H$\delta$ in absorption are classified as Post-starburst galaxies (PSB). The detailed description and numerical limits on each of the above selection criteria are discussed in \citet{poggianti09,fritz14}, and \citet{pacc17}. 

\subsection{Dynamic environment of A3376}
\label{denv}

The post-merger environment of A3376 exhibits complexity beyond the relaxed cluster environment, making it a two-fold problem to recognise any environmental influence: a pre-merger relaxed cluster environment which the member galaxies have been experiencing all along, and the post-merger dynamic environment where the cluster galaxies are being reshuffled while encountering the heated disturbed ICM and the outward shock due to the merger. Assuming a spherically symmetric shock front to have passed through the ICM $\sim0.6$ Gyr back \citep{kale12,lijo15,urdampilleta18}, originating at the onset of merger with a very low impact parameter, the innermost region ($\leq 0.5$ Mpc) would have experienced the shocked environment for the longest time with the boundary of the shock, denoted by the diffuse radio relics, situated almost $\sim1$ Mpc from the centre. Hence, it is highly likely that the galaxies located well beyond 1 Mpc may have never experienced this outgoing shock. 
We thus quantify the post-merger environment of A3376 through concentric regions of projected radius of 0.5 Mpc (0.25 R$_{200}$), 1 Mpc (0.5 R$_{200}$) and 1.5 Mpc (0.75 R$_{200}$), measured from the centre of the cluster system \footnote{Taking advantage of the geometry of the cluster merger and the low impact parameter, the centre of the A3376 cluster system is defined as the midpoint of the line joining the two BCGs (RA=6$^h$01$^m$25$^s$.00, Dec=-39$^\circ$59$^\prime$51$^{\prime \prime}$.00), with the merger axis being along this line. This definition remains consistent to the entire environmental analysis presented in this paper, unless mentioned otherwise.}. Note that for the purposes of this study we do not treat the A3376N group members independently but as a part of the whole A3376 merger system.

\section{Galaxy sample}
\label{s-galaxysample}

Our main galaxy sample comprises member galaxies of A3376 which have a spectral type and a morphology determined from the $V$-band images. Our base sample thus come out to be composed by 206 galaxies out of a total of 251 cluster members.

\begin{table*}

  \begin{center}
    \renewcommand{\arraystretch}{1.5}
    \centering
    \caption{The base cluster galaxy sample utilised in this paper in each spectral class (Section \ref{sp-class}), grouped according to their broad morphologies as mentioned in Section \ref{ss-morph}, and corrected for spectroscopic incompleteness \citep[mentioned in parantheses,][]{moretti17}.}\label{table1}
    \begin{tabular}{lcccc}

    \centering
    \textbf{Morphology} & Passive & Star-forming & Post-starburst & EML \\
  
    ($V$-band) & k & e(c)  & k+a/a+k & e(a)/e(b) \\
    \hline
    \hline
    Ellipticals (E) & 35 (41) & 6 (7) & 3 (3) & 0 (0)\\ 
    Lenticulars (S0) & 59 (70) & 13 (16) & 9 (11) & 3 (3) \\
    Spirals (S) & 43 (53) & 44 (53) & 8 (10) & 18 (22) \\
    \hline
  
    \end{tabular}
  \end{center}
\end{table*}

Throughout this paper, we define `star-forming' galaxies as those with spectral type e(c) and SFR$>0$ while `passive' galaxies as those having k-type spectra. Table~\ref{table1} enumerates the galaxies of each spectral type in the global cluster environment of A3376. 12 spiral galaxies from A3376 (not included in Table~\ref{table1}), originally classified as PSB or passive galaxies, were found to display emission lines and hence were reassigned an ambiguous e(a)/e(b)/e(c) EML galaxy status \citep{pacc17}. Upon carefully checking the galaxy spectra, colours, and computing aperture-limited SFR using H$\alpha$ emission line, we confirm that 11 of these 12 galaxies indeed are e(c)-type galaxies albeit with redder colours and minimal SFR. For the majority of them, the SFR estimates are below the detection threshold of OmegaWINGS\footnote{The sSFR detection threshold for OmegaWINGS Survey = $10^{-12.5}$/yr \citep{pacc16}}. Therefore, we decided to not include them in our sample of star-forming spiral galaxies owing to their sSFRs being lower by at least a factor of 10 than the rest of the spirals comprising our `star-forming' sample.

With the aim of comparing galaxy populations in A3376 to those in non-merging `relaxed' clusters at similar epoch, we obtain the ancillary data for field and relaxed clusters from the OmegaWINGS dataset. We confirmed the dynamical state of the clusters to be `relaxed' or `disturbed' based on qualitative visual classification of their X-ray morphology obtained from archival data. We then verified the existence of robust spectroscopic data and morphology information for the member galaxies of clusters identified as dynamically `relaxed', and select those clusters which lie within the redshift slice about the $z_{cl}$ of A3376 which we choose to be $0.03\leqslant z<0.06$. Our final cluster reference sample thus comprises 4 relaxed clusters and the complementary field. Table \ref{table2} enlists the basic properties of these clusters, and shows that all our selected relaxed clusters have similar masses and are within narrow redshift slice thus removing any bias as introduced by $z$-evolution of clusters. This gives us 855 cluster galaxies and 83 field galaxies in total.

\begin{table}
	\centering
	\caption{The control sample of relaxed clusters taken from the WINGS dataset. The columns indicate the name of the cluster, cluster $z$, cluster velocity dispersion, virial radius and cluster mass within the virial radius.}
	\label{table2}
	\begin{tabular}{ccccc} 
		\hline
		Cluster & $z_{cl}$ & $\sigma$ & $R_{200}$ & $M_{200}$\\
			& 	& km/s    & Mpc & 10$^{15} M_{\odot}$ \\
		\hline
		A151 & 0.053 & 738 & 1.8 & 0.67\\
		A1631a & 0.046 & 760 & 1.8 & 0.74\\
		A193 & 0.048 & 764 & 1.8 & 0.75\\
		A957 & 0.045 & 640 & 1.5 & 0.44\\
		\hline
	\end{tabular}
\end{table}

\section{Results}
\label{results}
\subsection{Galaxy populations in A3376}
\label{s-gpop}

N-body SPH Simulations of A3376 \citep{machado13} show that A3376 E and A3376 W clusters went through each other with very low impact parameter, and presently are travelling farther apart. We would thus expect galaxies of different morphology and spectral types to continue redistributing at such an early epoch after the merging activity. 

\subsubsection{Spatial distribution of spectral classes}
\label{spec-dist}
\begin{figure*}
	\includegraphics[width=\textwidth]{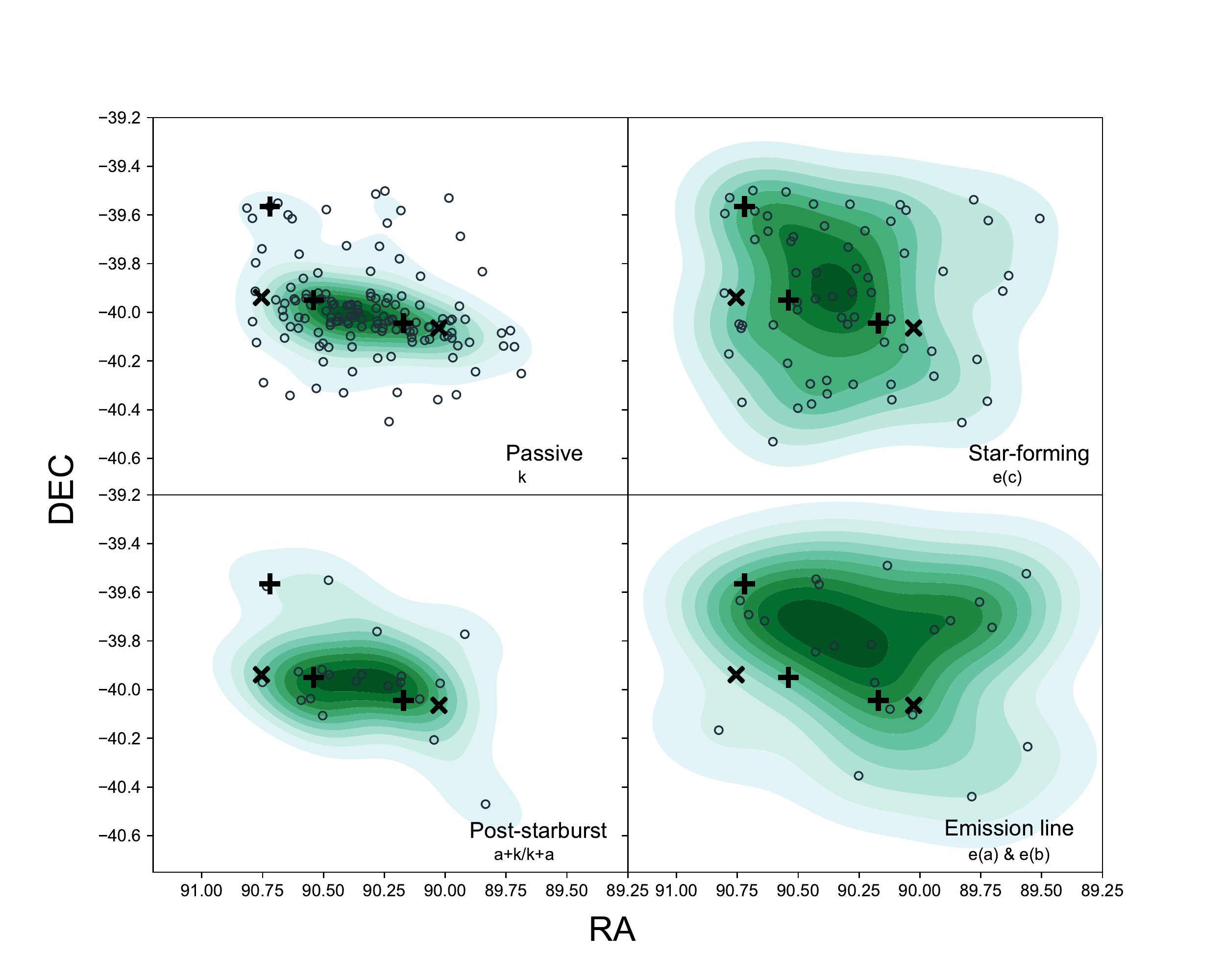}
    \caption{Projected distribution of cluster galaxies in A3376 according to their spectral classes derived from OmegaWINGS spectra. The plus signs denote the positions of BCG E, BCG W and BCG N while the the green contours show the projected density of galaxies, corrected by spectroscopic completeness, which are marked by open black circles.  The relative positions of the brightest points of radio relics are given by the crosses thus marking the position of the shock front. The passive and post-starburst galaxies are more concentrated about the merger axis, i.e. the line joining the two BCGs, while the star-forming galaxies appear to have a wide spatial distribution. The emission-line galaxies however show an interesting asymmetry in their distribution skewed towards the north of the cluster. }
    \label{spec_dist}
\end{figure*}

Figure~\ref{spec_dist} shows the spatial distribution of galaxies in each of the aforementioned spectral classes within A3376 system, with the green contours showing the observed spatial density. The galaxy populations, in general, show a large degree of redistribution symmetric about the merger axis, except perhaps for the emission-line galaxies. The passive galaxies in A3376, mostly expected to reside in dense environments around the BCGs, are uniformly distributed between the two BCGs. A wider distribution is observed in the star-forming galaxies, skewed more towards the northern Group, and having higher line-of-sight velocities. This is not surprising since they are expected to be more common in low-density cluster environments. 

Several studies have found that cluster PSBs avoid dense cluster cores \citep{poggianti09-2,rudnick17}. We find a hint of symmetry in the projected distribution of PSBs though the numbers are less to robustly support this. However, further spatial and line-of-sight velocity analyses are required to ascertain their location in A3376, which are presented later in Section~\ref{psb}. Similar to star-forming galaxies, other emission-line cluster galaxies display a highly asymmetric distribution skewed northwards of the system. This may partly be attributed to one of the clusters having a higher incidence of emission-line galaxies which are getting redistributed. The nature of these emission-line galaxies, i.e whether star-forming or due to active galactic nuclei, will be presented in upcoming OmegaWINGS data release of the extended emission-line catalog (Radovich et al., in prep). Altogether, the projected spatial distribution of EML galaxies seems to be coherent with findings of \citet{pacc17} where they find a higher incidence of EML galaxies at higher cluster-centric radii in all OmegaWINGS clusters. Unlike the fraction of PSBs which appear to be unchanged as a function of cluster-centric radius \citep{pacc17}, the PSBs in A3376 tend to be prevalent in the central regions of the cluster mimicking the distribution of passive galaxies. In general, while symmetry is observed around the axis of the merger we do not find any clear evidence of preferential distribution of either spectral classes near or across the radio relics. Nonetheless, the emission-line galaxies do demand a targeted analysis which will be presented in upcoming papers of this series. Apart from a small overdensity of passive galaxies north of BCG E, we do not find a clear indication of the presence of the A3376 N group.

\subsubsection{Spatial distribution of Hubble classes}
\label{morph-dist}

\begin{figure*}
	\includegraphics[width=\textwidth]{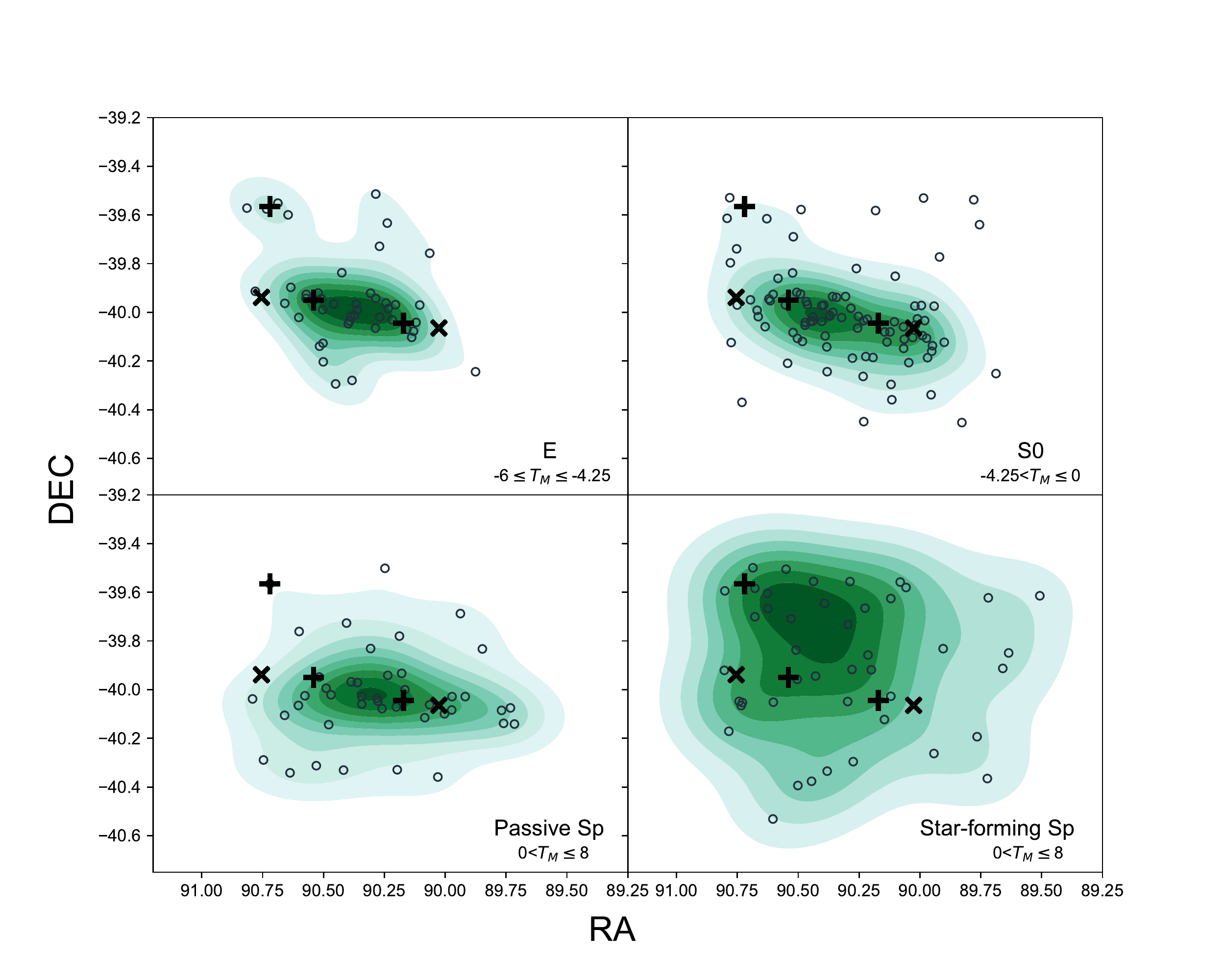}
    \caption{Projected distribution of cluster galaxies in A3376 according to their broad $V-$band morphologies, given by the \textsc{morphot} morphology type ($T_M$) in each panel. The ellipticals and lenticulars show dense spatial distribution about the merger axis. As evident from their density contours, the passive spirals show similar concentration as the early-type galaxies while regular star-forming spirals appear to be more wide spread.}
    \label{morph_dist}
\end{figure*}

We examine the spatial distribution of Ellipticals (E), Lenticulars (S0) and spirals (Sp) in A3376, displayed in Figure \ref{morph_dist}. As expected, we observe the ellipticals to lie in the densest region around the two BCGs and in between. This elongated distribution of Es is not surprising because previous studies \citep{machado13} investigating the merger dynamics of A3376 conclude that the most massive cluster A3376 W got completely disrupted during the pericentric passage and hence could result in the elongated distribution of the centrally located galaxy populations. Although the lenticulars show spatial concentration similar to ellipticals, they appear to have wider spatial distribution in the outskirts of the cluster system. Spiral galaxies are observed to be more abundant in less dense regions \citep[for e.g.,][]{fasano15}. Moreover, several studies have shown that a fraction of cluster spirals most often are a population in transition, thus supporting the reported quenching of star formation in the cluster environment and the consequent structural transformation \citep{wolf09,bosch13,kelkar17}. We account for this by investigating the spatial distribution of spiral galaxies (lower panels of Figure~\ref{morph_dist}), according to their passivity (k-type) and star-forming activity (spectral type e(c) and SFR $>0$). It is interesting to note that the spatial distribution of passive spirals mimics that of lenticulars and ellipticals while the star-forming spirals are more dispersed across the cluster. The likely reason for this dichotomy is that we may be observing a subset of spiral galaxies which are already quenched and getting accumulated on the cluster core \citep{kelkar19}. 

\begin{figure}
	\includegraphics[width=\columnwidth]{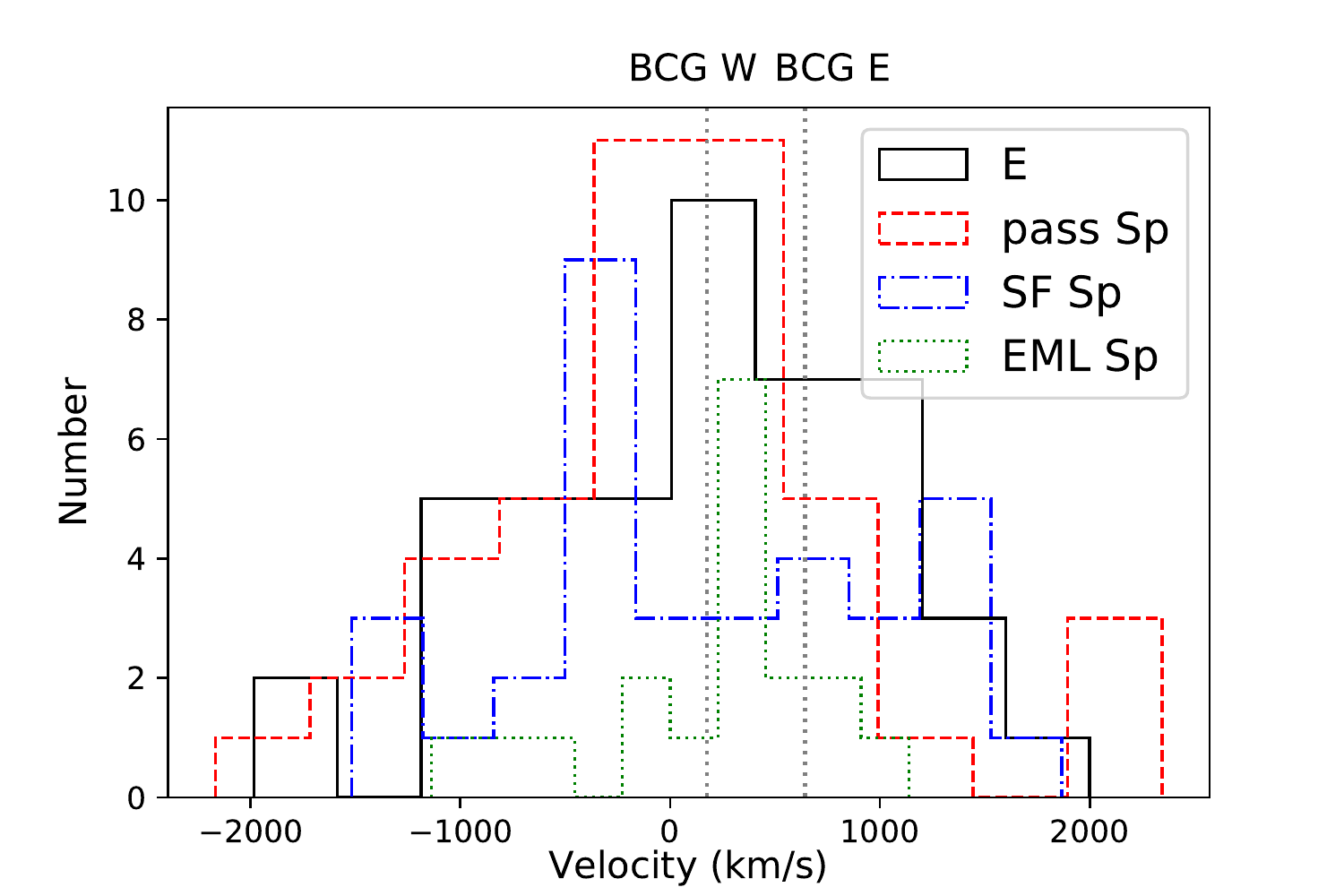}
    \caption{Line-of-sight velocity distribution for cluster ellipticals, passive spirals and star-forming spirals from A3376, computed with respect to $z_{cl}$. The line-of-sight velocities for passive spirals imitate that of ellipticals while the velocities for star-forming spirals are more widely distributed. A small peak of star-forming spirals about the velocity of BCG E correspond to the subset of massive blue spirals discovered on the $(B-V)_{o}$ colour--magnitude relation of A3376, and located in the cluster center (Section~\ref{sf_a3376}).}
    \label{vlos}
\end{figure}

We also look at line-of-sight velocities, measured with respect to the $z_{cl}$ of A3376, to gauge the true distribution of spiral galaxies. Figure~\ref{vlos} shows the line-of-sight velocity histograms for the ellipticals, passive and star-forming spirals, and EML galaxies in our cluster. Elliptical galaxies indeed occupy the densest regions (corroborated by the low relative velocity of ellipticals) of the cluster, mostly around each BCG. The asymmetry in the velocity distribution is likely due to unequal cluster masses, with BCG E located in the least massive but compact A3376 E. We also see a significant peak of passive spirals specifically around BCG W. Combining with the spatial distribution given in Figure \ref{morph_dist}, we can be assured that these passive spirals indeed are located in the central regions of the cluster, preferentially around BCG W. Although their line-of-sight velocity distribution is broad, such an asymmetric distribution of passive spirals is best explained by the fact that A3376 W is a massive cluster and hence would contain more spirals with their star formation already suppressed before they transform morphologically. The majority of the star-forming spirals, on the other hand, are located in the cluster periphery owing to the suppression of galaxies' star-forming activity due to the dense cluster environment. Surprisingly, a small subset of star-forming spiral galaxies appear to be situated in the dense regions of the cluster where one would expect minimal to none ongoing star formation in galaxies. These galaxies will be revisited in the next section through a detailed analysis of the dynamic environment the galaxies experience post merger, and the resultant effect on the star formation histories of the galaxies. With most of the EML galaxies having spiral morphologies and wider spatial distribution (Figure \ref{spec_dist}), we note that they have line-of-sight velocities indicating that they are mostly peripheral/infalling galaxies.   

\subsection{Star formation in A3376 }
\label{sf_a3376}

As mentioned before, massive dynamical events like cluster mergers often leave their imprints on the ICM and global structure of the systems. However, their precise role in shaping the inherent galaxy populations remains unclear. Our analysis becomes one of a kind targeting the galaxies which have experienced such post-merger dynamic environment and may have had alterations in their star formation activity. Our preliminary phase--space analysis, i.e. combination of spatial (Figure \ref{morph_dist}) and line-of-sight velocity distribution (Figure \ref{vlos}), brings to light interesting trends in the spiral galaxy population when observed according to their passivity or star formation. While it may very well be probable that the global cluster environment is acting on spiral galaxies, there is a likelihood of cluster merger-driven phenomenon at play, causing changes in star formation activity of galaxies.

\begin{figure}
	\includegraphics[width=0.9\columnwidth]{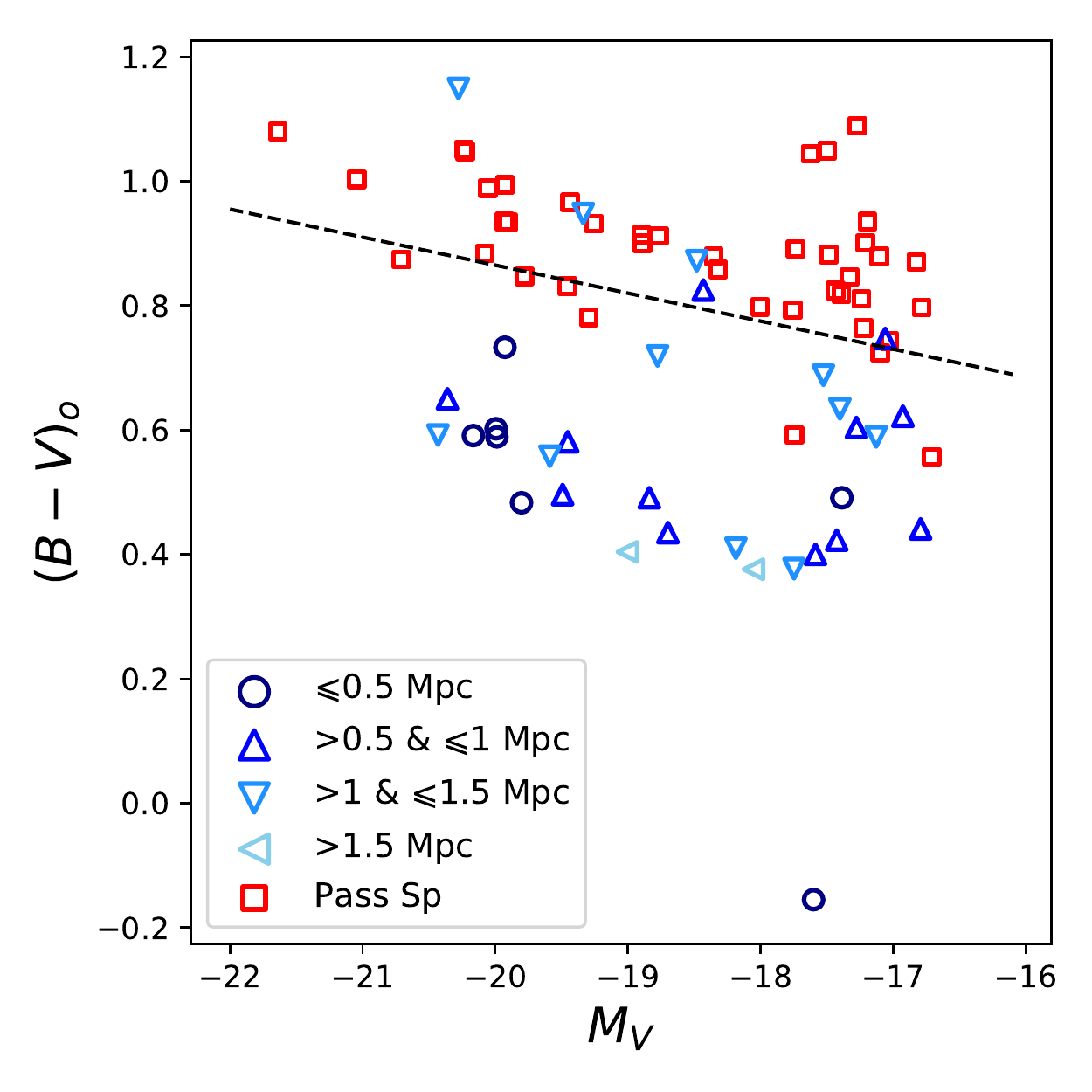}
    \caption{The rest-frame ($B-V$) vs $M_{V}$ colour--magnitude relation for spirals in A3376. Star-forming spirals are denoted by circles, and triangles colour-coded according to their distance from the cluster centre while passive spirals are collectively displayed in red squares. The dashed line represents the colour--magnitude red sequence given by \citet{valentinuzzi10-2}. We report a small subset of massive spirals showing bluer colours and located in close to the cluster centre (dark blue circles). This result implies that these galaxies are actively forming stars despite being in harsh environment. The passive spirals in contrast are majorly on the red-sequence affirming that they indeed are not forming stars and likely represent a quenched population of spirals yet to undergo morphological change. }
    \label{cmr1}
\end{figure}

Utilising the rest-frame $B-$ and $V-$ band magnitudes from \citet{gull15}, we first inspect the $B-V$ vs $V$ colour--magnitude relation (CMR, Figure~\ref{cmr1}) of our member galaxies as a first-order diagnostic to identify plausible effects of cluster merger on them. Exploited as a photometric indicator of current star formation activity, any galaxy transiting from a star-forming phase to a passive one could be identified on this CMR as population making the jump from the star-forming `blue cloud' to the passive `red-sequence', with a possibility of catching them in the intermediate `green valley'. We use the colour--magnitude red sequence given by \citet{valentinuzzi10-2} for the WINGS sample, and later adopted for OmegaWINGS sample by \citet{pacc17}. The galaxies are allocated to the red-sequence if their colour lies above 
\begin{equation}
 (B-V)_{o}=-0.045\times M_{V}-0.035,
\end{equation}
where $M_{V}$ is the absolute $V-$band magnitude.

As a whole, our sample appears to follow the expected trends on the CMR, when selected according to their spectral types. Distinct spectral classes populate specific regions of CMR with the passive galaxies located on the red sequence while the star-forming and emission-line galaxies concentrated in the bluer regions of CMR. Such a distinct bifurcation, however, gets diluted when viewed across different morphology classes. 

\subsubsection{Passive and blue spirals in cluster centre: pre- and post-merger signatures?}

With ellipticals and lenticulars already in place on the red sequence, we note a significant fraction of spiral galaxies having redder colours and already populating the red-sequence. This potentially supports the scenario where cluster spirals are quenched on timescales smaller than their structural transformation in high density environment \citep[e.g.][]{wolf09,kelkar17}.

We focus on the regular star-forming spirals in A3376, and report a significant spread in the distribution of blue star-forming spirals especially towards the brighter end (i.e. higher $V-$band magnitudes) of the CMR. We introduce the post-merger dynamic environment by analysing the rest-frame CMR for spirals in A3376 according to their spectral type and location in the cluster. We focus primarily on the star-forming and passive spirals according to their projected distances from the cluster centre shown with different colors/symbols in Figure~\ref{cmr1}), with Eq 1 as the fitted red sequence.

\begin{figure}
	\includegraphics[width=0.9\columnwidth]{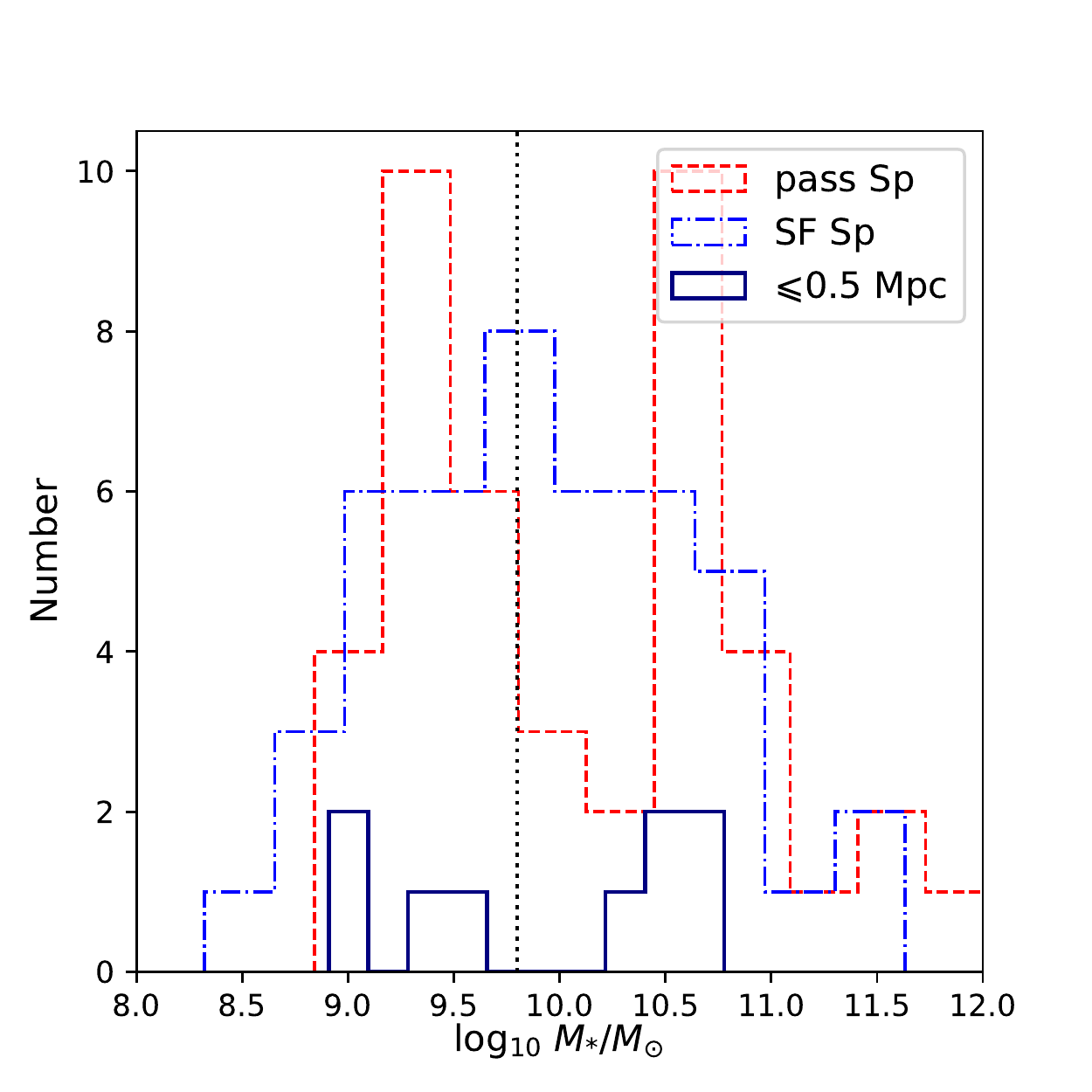}
    \caption{Stellar mass distribution of passive (dashed red) and star-forming (dash-dot blue) cluster spirals in A3376. The passive spirals denote interesting bi-modal distribution of stellar masses about the mean stellar mass (vertical dotted black line) of the spiral galaxy sample. However majority of the star-forming spirals within the central $\leqslant$0.5 Mpc of A3376 (solid blue) have stellar masses higher than the mean stellar mass.}
    \label{mstar-hist}
\end{figure}
We further discover a subpopulation of blue spiral galaxies located in the innermost region of our cluster (dark blue circles in Figure~\ref{cmr1}) and having bluer colours. Figures \ref{morph_dist} and \ref{vlos} confirms that $>$50\% of these blue star-forming spirals have line-of-sight velocities within $\pm$250 km/s to that of BCG E (Figure \ref{vlos}). This observation has important connotations to our understanding of star-forming galaxies in cluster environments. Along with the possibility of being in the central region of the cluster between the two BCGs, most of these spiral galaxies are massive (Log $(M_{*}/M_{\odot})>10$, Figure~\ref{mstar-hist}) and their location on the CMR suggest that they are actively forming stars while inhabiting the shock-heated ICM of the post-merger system.

\begin{figure}
	\includegraphics[width=\columnwidth]{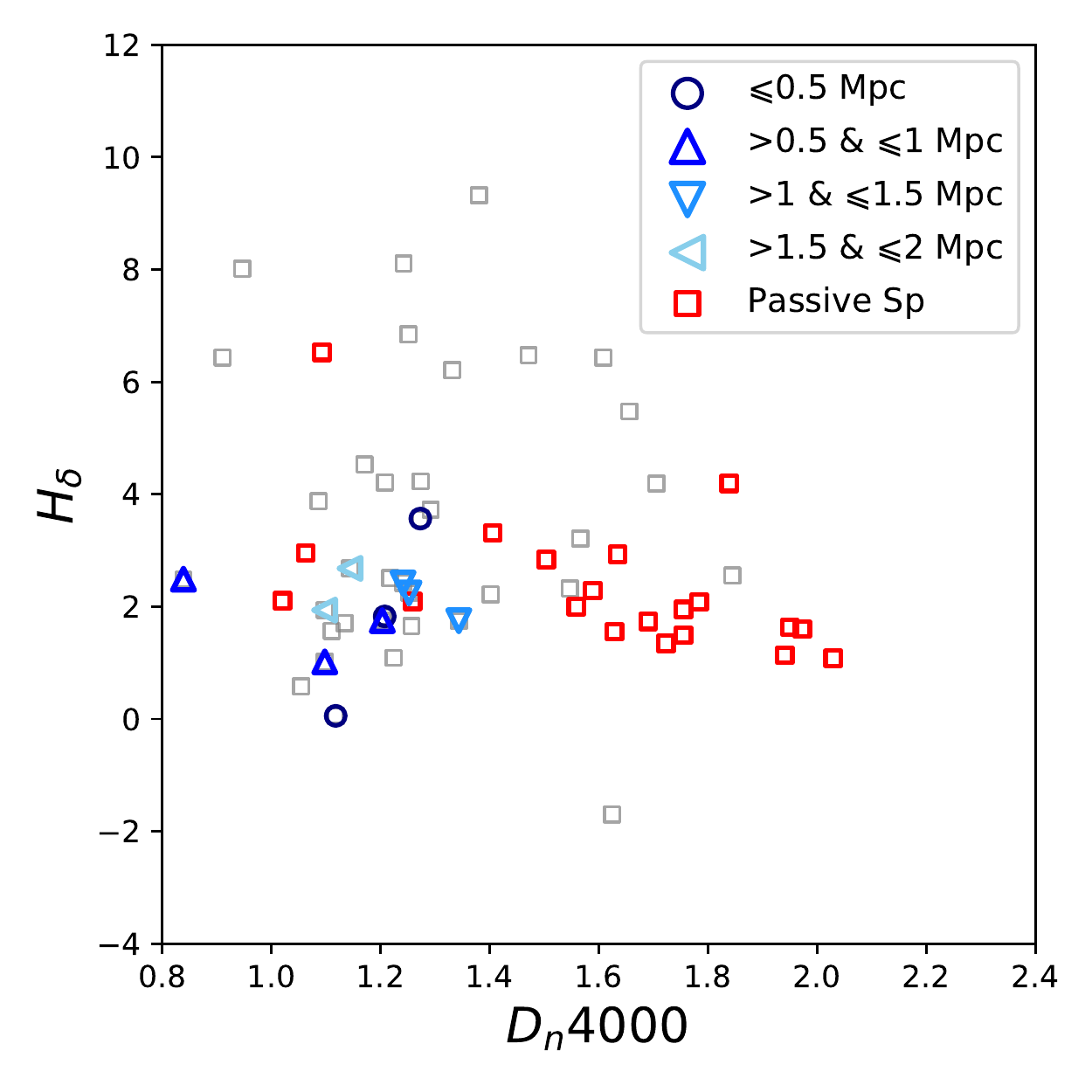}
    \caption{Balmer absorption vs $D_{n}4000$ break strength for spiral galaxies in A3376 Continuing the symbol and colour assignment from Figure~\ref{cmr1}, we highlight the star-forming and passive spirals in A3376 while rest of the spirals are denoted by gray points. As expected, the star-forming spirals have younger stellar populations as compared to passive spirals. However, there does not seem to be clear trend with respect to distance from cluster centre although low numbers restrict us from concluding anything. }
    \label{sfh}
\end{figure}

We explore the star formation history (SFH) of these galaxies using the 4000 \AA{} break strength (D$_{n}4000$) and H$\delta$ equivalent width (EW) in absorption. The choice of these indices lies in the ability of H$\delta$ to be indicative of time since the last episode of major star formation. This becomes crucial in our studies as H$\delta$ absorption EW peaks around 0.5-1 Gyr since the star-forming activity, which coincides with the timescale of the merger of A3376. The D$_{n}4000$ on the other hand gives a good estimate of the luminosity-weighted ages of the stellar populations within the galaxies. Hence, any considerable star-forming activity within past $\sim1$ Gyr should be able to be detected through the combination of these indices. Figure \ref{sfh} shows the H$\delta$--D$_{n}4000$ plot for a fraction of our cluster spirals, for which D$_{n}4000$ and H$\delta$ in absorption were identified and measured by \textsc{sinopsis}, as a function of their distance from the cluster centre. As expected, we note clear segregation between passive spirals and star-forming spirals with most of the passive spirals having older stellar populations i.e. higher D$_{n}4000$. The star-forming spirals, however, do not show any difference in their star formation history as a function of the dynamic environment. The low H$\delta$ EW, especially for three of the five massive blue galaxies within $\leq$ 0.5 Mpc, further supports a lack of change in the star formation history of these galaxies.

\begin{figure}
	\includegraphics[width=\columnwidth]{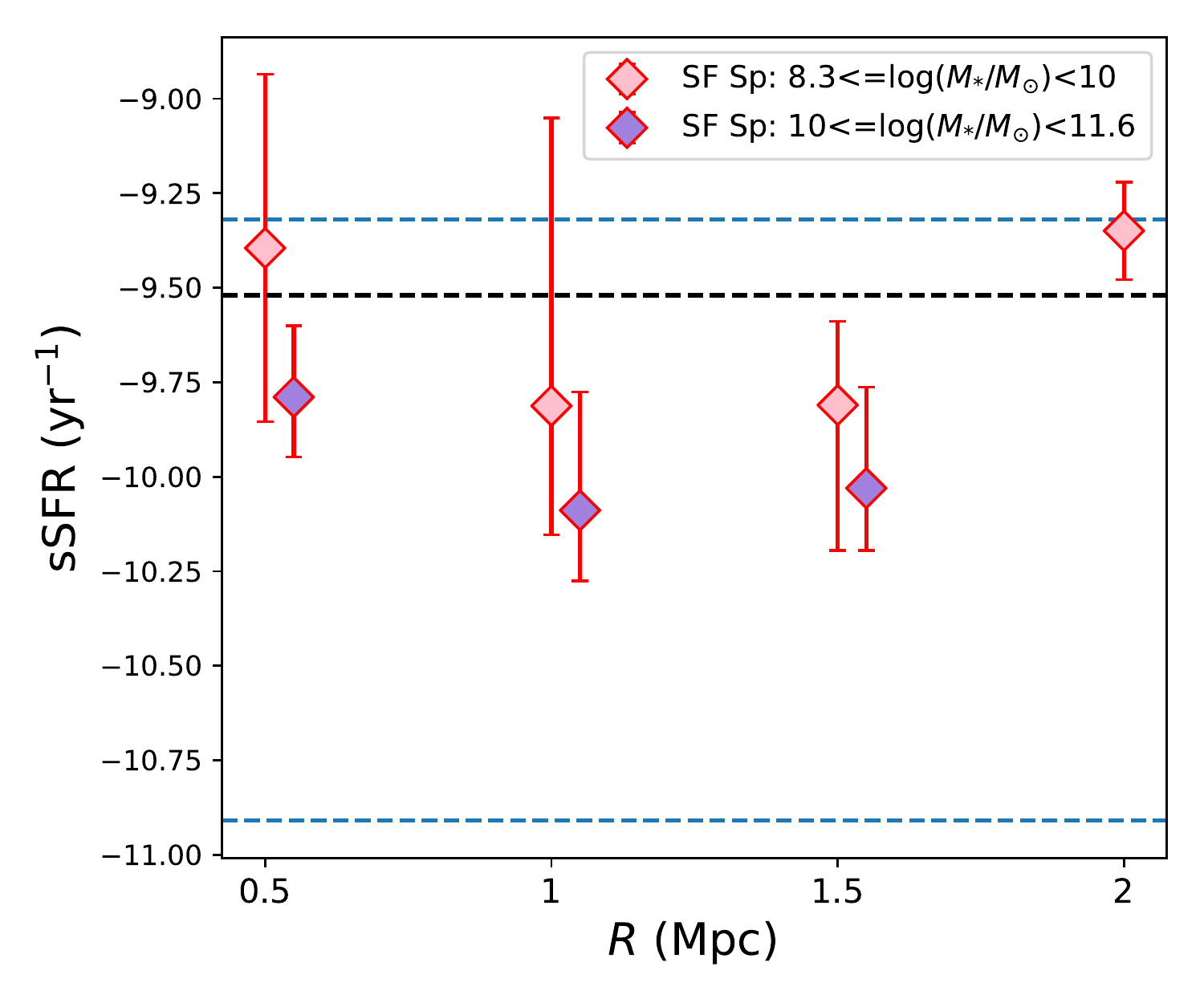}
	\caption{Median specific star formation rate (sSFR) in each cluster-centric radius (R) bin for high-, and low-mass star-forming spirals. The gray dashed lines indicate the field sSFR values for the minimum and maximum stellar mass in our sample while the black dashed line denote the same for the mean stellar mass of our sample. Overall the star-forming spirals, especially the high-mass spirals seem to have sSFR comparable to the field galaxies though no particular trend as function of cluster-centric radius is observed. Note that no high-mass star-forming spirals were found beyond $R>$1.5 Mpc.}
    \label{ssfr-R}
\end{figure}

Finally, we analyze the variation of the specific star formation rate of cluster spirals, broadly binned according to their stellar masses. Figure \ref{ssfr-R} shows the median specific star formation rate (sSFR), with error bars denoting the limits, in each concentric areas around the centre of A3376 for low- and high-mass bins. We observe that the sSFR -- environment relation seem to hold true \citep[e.g.][]{lagana18} even for recent post-merger systems like A3376, where sSFR does not seem to depend on the cluster centric radius. We overplot the sSFR derived for field galaxies from the control cluster sample (Table \ref{table2}) computed for the lowest, highest, and the mean stellar mass in our sample (dashed lines), though it should be noted that our sample is incomplete for the low-stellar mass bin \footnote{In order to be consistent with previous WINGS literature and the ongoing data release of OmegaWINGS, we also adopt the mass completeness limit of Log $(M_{*}/M_{\odot})=9.8$ \citep{pacc17}}. Though the average sSFR appears to remain unchanged as a function of cluster radius, we observe that the overall sSFR of high- and low-mass galaxies to be within the field reference at all cluster-centric radii. Moreover, we observe a marginal spread in sSFR for low-mass galaxies especially towards within $<$ 1 Mpc radius suggesting more low-mass spirals to have higher star formation than a the average field population. Contrasting with the expected trends showing suppressed star formation, we infer that the star-forming spirals in A3376 appear to have star formation comparable to field galaxies of similar masses. We however caution that the low numbers of low-mass galaxies in the central region of A3376 and the incompleteness associated with low-mass galaxies restricts us from deducing the trend definitively. Furthermore, the spatial and line-of-sight velocity analysis discussed in the previous section suggests that there is a high possibility that most of these low-mass spirals might be located in cluster outskirts observed in projection thereby introducing the spread in sSFR thus observed. 

In a nutshell, our primary analysis using positional information, and SFH inferred from spectroscopic and photometric data demonstrates that a few massive spiral galaxies in A3376 are undergoing star formation despite residing in a supposedly hostile environment, which recently underwent a dynamical large-scale interaction through cluster merger. It is plausible that these massive spirals may have been in the periphery of their parent clusters before the merger, and could be brought in the path of merging. Our results thus hint at the possibility that the merger shock could be responsible for the continued star formation in the high-mass cluster spirals as it passes through the galaxies. However, a dedicated study is needed of other galaxy populations in order to place this observation in the overall picture of how post-merger shock can affect the star formation in cluster galaxies.

\subsubsection{Post-starburst galaxies in A3376}
\label{psb}
 
Several studies have explored the origins of PSBs in clusters which link the `k+a' signatures in galaxy spectra to the interaction with hot ICM in cluster environment \citep{poggianti04, poggianti09-2}, though contrasting observations at low-$z$ suggest that PSBs prefer low-density environment \citep{goto05}. Interestingly, the role of PSBs as tracers of ICM interactions on galaxies' star formation in the light of larger-scale cluster mergers is largely unexplored. Simulation studies like \citet{bekki10}, however propose that the increase in the ICM pressure during a cluster merging event can trigger star formation in gas-rich cluster spirals which could be transformed into post-starburst galaxies. 

These `k+a' signatures arise when a continuous star formation encounters an abrupt suppression or when a major star-forming episode is recently truncated. The PSBs thus offer a very unique timestamp to recent SFH of galaxies since the spectral signatures are characteristic to newborn A-type stars born within the last $1$ Gyr or so, with the features visible in the galaxy spectra only till $\sim1-1.5$ Gyr since the truncation of star formation. This puts the merger system of A3376 in a crucial position wherein the timescale of the pericentric passage of the two merging clusters in A3376 coincides very well with that of the k+a signatures in galaxy spectra. We thus present the first ever census of PSBs in a post-merger cluster system and investigate the probable influence of the large-scale merging activity on the possible origins of k+a galaxies in A3376.

\begin{figure}
	\includegraphics[width=\columnwidth]{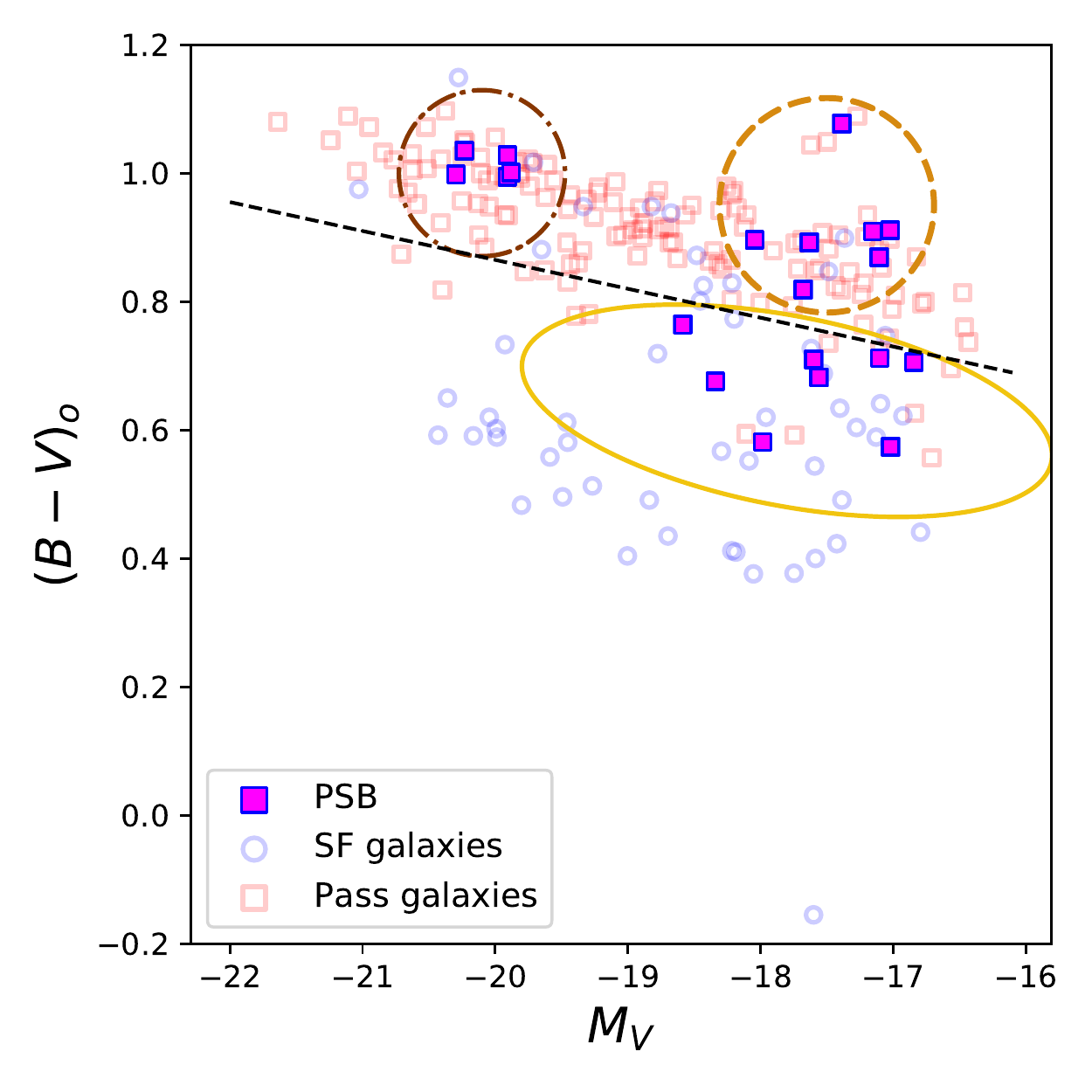}
    \caption{Rest-frame absolute $B-V$ vs $M_{V}$ colour--magnitude relation for PSBs in A3376. The overplotted dashed line represents the red sequence fit according to \citet{pacc17}. The PSBs in A3376 appear to be well seprated in three broad subpopulations thus hinting at different ages of PSBs in A3376- `Bright' PSBs are marked by the dash-dot circle, while the dashed circle denotes the `Faint' PSBs with the low-mass blue PSBs shown by the solid circle.}
    \label{cmr-psb}
\end{figure}

\begin{figure*}
	\includegraphics[width=1.3\columnwidth]{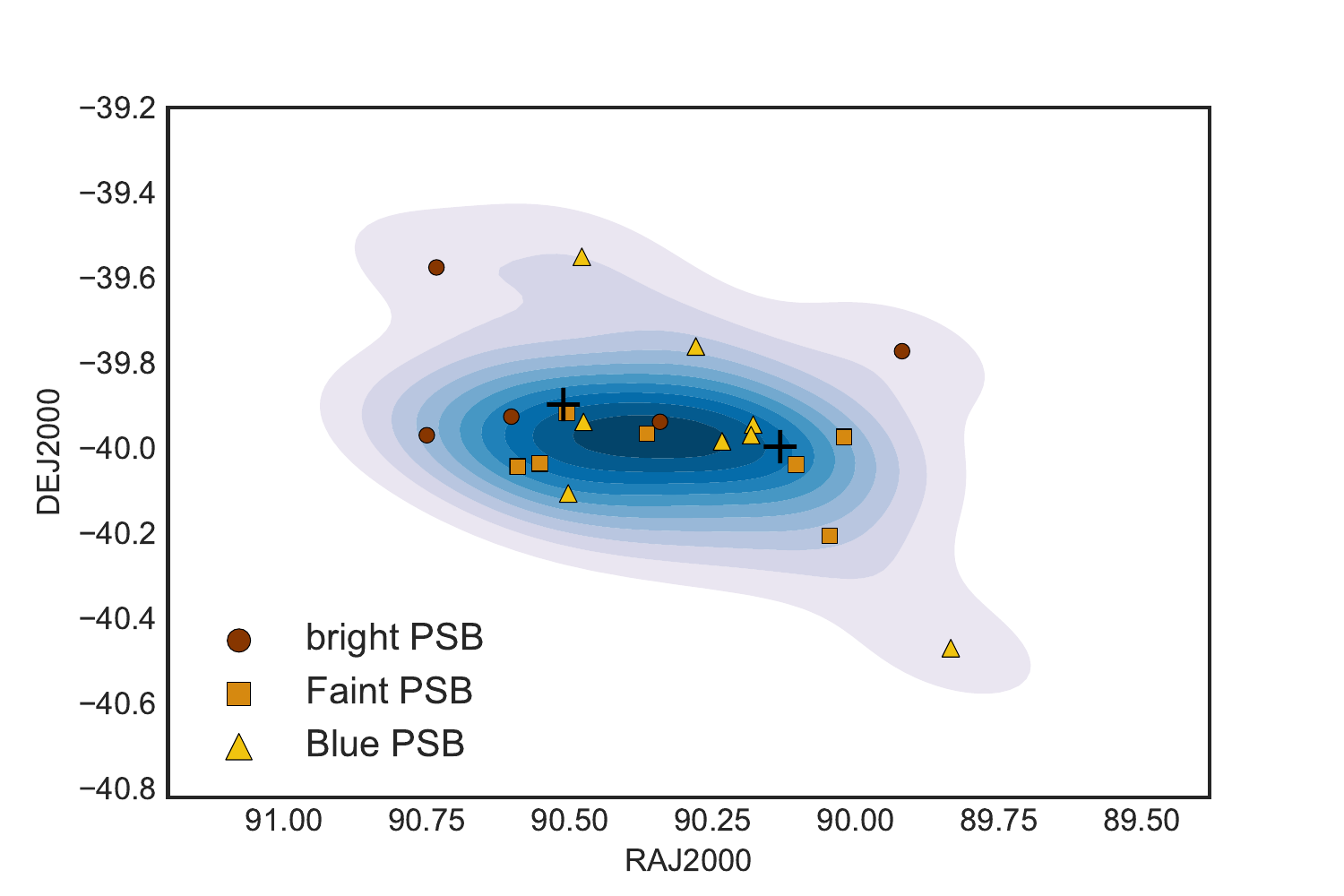}
    \caption{The projected spatial distribution of PSBs as a function of location on the CMR. Most of the blue PSBs tend to be concentrated along the axis of merger between the two BCGs although similar trends in bright or faint PSBs are unclear. }
    \label{psb_sea}
\end{figure*}

We construct a sample of PSBs by selecting galaxies displaying `k+a'/`a+k' spectra, as discussed Section \ref{sp-class} (See also Table ~\ref{table1}). As with the cluster spirals, we introduce Figure \ref{cmr-psb} showing the distribution of cluster PSBs on the absolute rest-frame $B-V$ colour--magnitude relation. Surprisingly we discover that the PSBs in A3376 appear to populate different regions of the CMR and hence must be different subpopulations. Previous studies with the full OmegaWINGS cluster sample revealed that the incidence of PSBs in galaxy clusters strongly depend on the dynamic state of the cluster, with relaxed clusters having a relatively higher fraction of PSBs than disturbed ones \citep{pacc17}. Although our analysis is limited to only one cluster system, the segregation of PSBs on the CMR suggest that the population of cluster PSBs are far more complex especially in merging clusters. With the known general picture of galaxy dichotomy on CMR, one would expect the PSBs to lie in the transition region between the red sequence and the blue cloud. A segregation of PSBs on the CMR therefore implies that the PSBs may have different formation histories and thus different ages. We divide the PSBs into the three main subgroups as observed on the CMR and use the following criteria-

\begin{align*}
(B-V)_{o}>-0.045\times M_{V}-0.035 \quad \& \quad M_{V}<-19 \\
(B-V)_{o}>-0.045\times M_{V}-0.035 \quad \& \quad M_{V}>-19 \\
(B-V)_{o}<=-0.045\times M_{V}-0.035, 
\end{align*}

to define the `bright', `faint' and the `blue' PSBs respectively. Their location on the CMR indicates that the bright and redder PSBs are massive galaxies as compared to the faint and blue ones. Morphologically, each of these groups of PSBs show interesting distribution: most of the bright PSBs are dominated by early-type disc morphology like the lenticulars/ S0s while the faint PSBs are found to equally display either spiral or lenticular morphology. The blue PSBs, on the other hand, are dominated by late-type spiral morphologies supporting the idea that these PSBs indeed are younger than the rest of the PSB subgroups. Structurally, a fraction of these blue PSBs shows visual asymmetries and signatures of disturbances likely associated with external mechanisms \citep[See also][]{kelkar17}. Their low numbers, however, limit us from inferring the plausible external triggers attributed to their blue colours and post-starburst phase. Also, the faint and blue PSBs span a very narrow range of magnitude ($M_{V}>-19$). Furthermore, the vast majority of the blue and faint galaxies have a stellar mass that is very close to the mass-completeness limit of the OmegaWINGS spectroscopic catalog, thus hampering a robust statistical analysis. Nevertheless, our data seem to suggest that these galaxies may belong to common subpopulation where the least massive/low luminosity galaxies undergo a strong post-starburst phase and age eventually to have redder colours. However, whether this is an observed trend in relaxed clusters \citep[See also Figure 2;][]{pacc17} or whether is characteristic to merging cluster systems is yet unclear. We may, however, be observing PSB populations already existing since before the cluster-merger happened ~ $\sim0.6$ Gyr ago (bright PSBs) while the onset of the merger itself triggering star formation in low-mass galaxies followed by a rapid quenching \citep{bekki10}. 

\begin{figure}
	\hspace{-0.3in}
	\includegraphics[width=1.1\columnwidth]{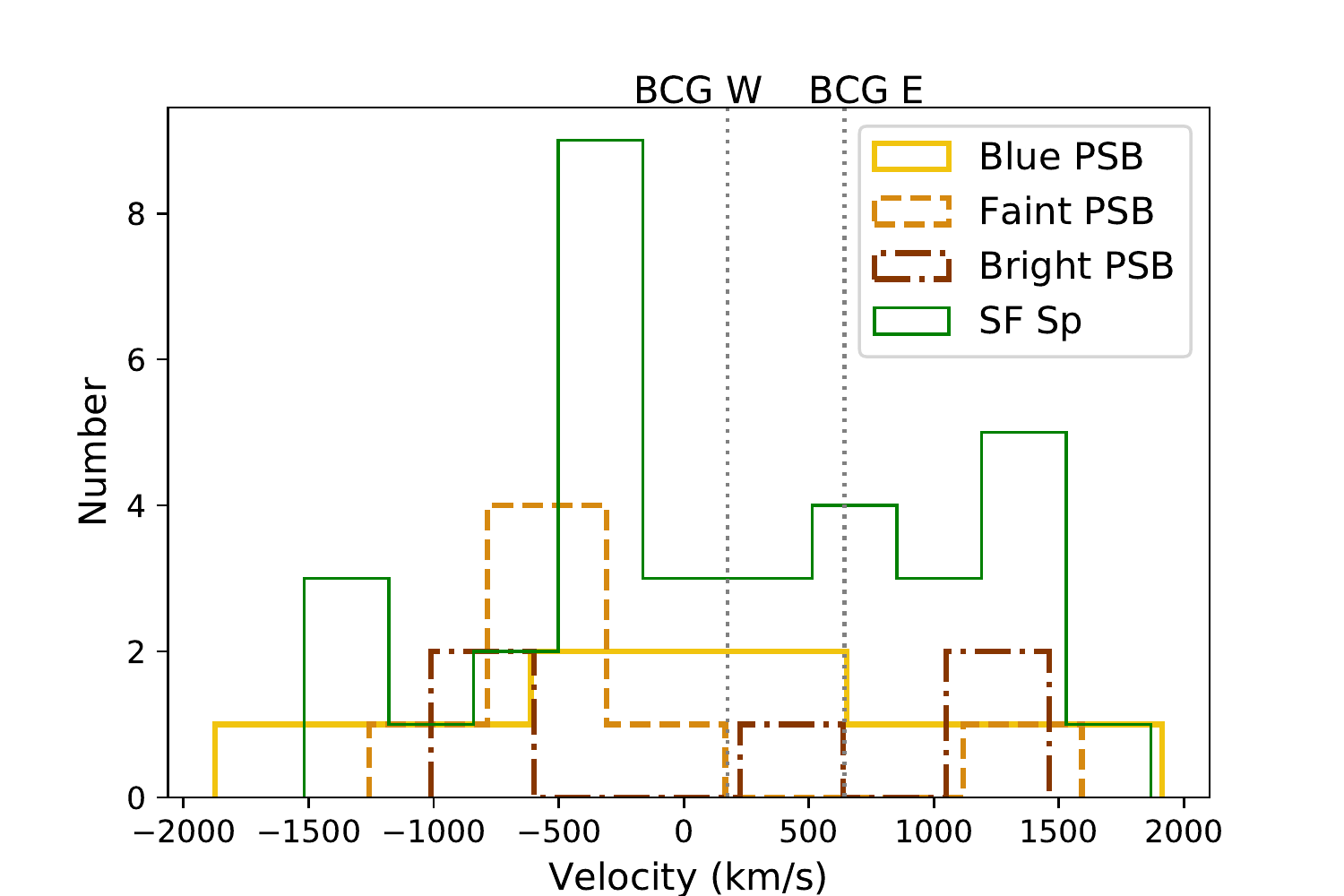}
    \caption{The line-of-sight velocity distribution for bright, faint and blue PSBs. With the velocity distribution for the star-forming spirals overplotted for reference, we notice that the blue PSBs have wide range of line-of-sight velocities. Together with the spatial distribution presented in Figure~\ref{psb_sea}, there is large uncertainty regarding the location of each these PSBs subclasses in A3376.}
    \label{psb_vlos}
\end{figure}

We see a trend in the SFH of these PSB galaxies that we think is not very much significant due to the low number statistics where the blue PSBs contain younger stellar populations with strong H$\delta$ in absorption and bright PSBs showing relatively higher D$_{n}4000$. We next check whether these PSB subgroups prefer different regions in A3376. Figure \ref{psb_sea} shows the spatial distribution of the bright, faint and blue PSBs with respect to the two BCGs in A3376. We observe all the PSBs to have an elongated distribution akin to that of passive galaxies. There is no difference however in the spatial distribution of bright, faint and blue PSBs. We also check the line-of-sight velocity distribution of PSBs with respect to the cluster centre (Figure ~\ref{psb_vlos}). Though the blue PSBs seem to be spatially concentrated in the central regions of the cluster, they seem to have the widest line-of-sight velocity distribution. We thus argue that the different ages observed in PSBs of A3376 are attributed to different formation ages and mechanisms. The bright red PSBs indicate a post-starburst phase initiated due to infall in the cluster likely before the merger. The blue younger PSBs, however, show a mixture of effects- recent infallers with higher line-of-sight velocities, and those galaxies whose k+a signatures arise from either the sudden surge in the ICM pressure at the onset of merger or the propagating shock-wave. These merger-driven blue PSBs thereafter underwent rapid quenching due to shock heated ICM \citep[See also][]{ma10}.   

Overall, our analysis of PSBs in A3376 suggests that blue, low-mass PSBs are strongly affected by the cluster merger activity. Any trends in the position of these PSBs within the cluster remain unclear. However, we robustly establish that A3376 has distinct PSB subpopulations, characterized by different ages, where the blue, low-mass tail is due to the youngest population.

\subsection{Global SF properties of cluster spirals in A3376 vs relaxed cluster spirals }
\label{sf-glob}

Recently, studies have shown that dynamically young massive merging cluster systems host an unexpected population of star-forming galaxies and active galaxies \citep{stroe14,owers12,ferrari05,pranger14} while they appear to be absent in equally massive systems but with different merger timescales \citep{stroe15a}. With a limited number of such systems studied with respect to the inherent galaxy populations, that too at higher $z$, the spread of dynamical disturbances in these systems is far too large to get a coherent idea of how evolving environment impacts the member galaxies and what observable signatures could be targeted to study the same. We thus step out to a global picture with the aim of identifying how different A3376 is with respect to its star-forming population when compared with relaxed clusters in a similar redshift range. We return to the sample of star-forming spirals from our A3376 and this time investigate the SFR of spiral galaxies as a function of galaxy stellar mass. 

\begin{figure}
	\includegraphics[width=\columnwidth]{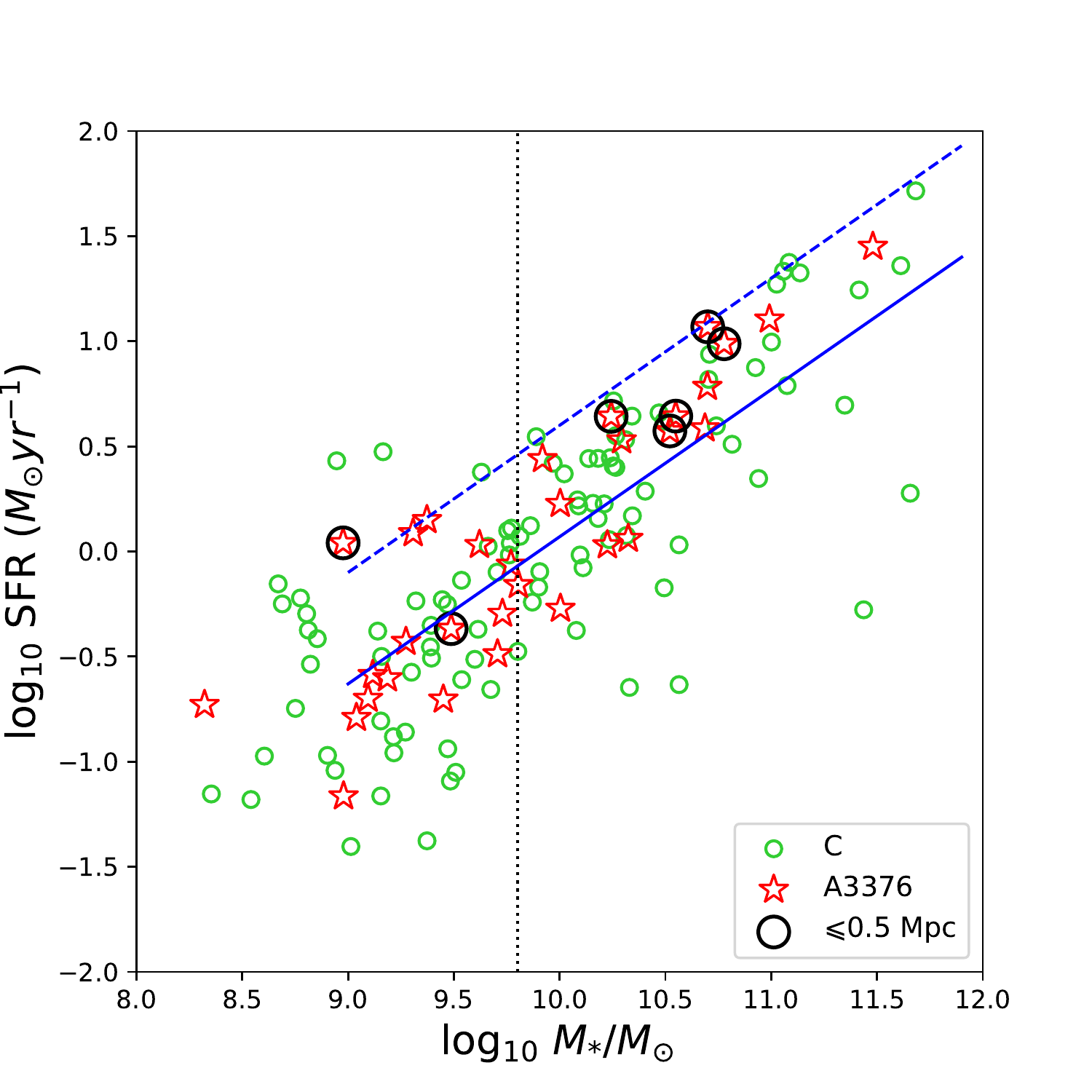}
    \caption{The SFR vs $M_{*}$ relation for star-forming spirals in A3376. The reference relaxed cluster sample (green circles), detailed in Table~\ref{table2}, and selected in the same way as the A3376 sample. The vertical dotted line denotes the mass-completeness limit for the OmegaWINGS data while the black circles indicate the blue star-forming spirals residing in the core of A3376 (from Figure \ref{cmr1}). The blue dashed line denote the best fit line for all the star-forming field galaxies in OmegaWINGS sample while the solid blue line indicate the 1.5$\sigma$ from the field best-fit line \citep{pacc16}. Comparison with OmegaWINGS survey data from \citep{pacc16} suggest that all the star-forming spiral galaxies in A3376 lie near the boundary line (blue dashed line) signifying that though they are not purely star forming (above the boundary). Thus all the star-forming spirals in A3376 have SFR similar to average star-forming cluster galaxies .}
    \label{sfr-mstar}
\end{figure}

We select the star-forming spirals from the control cluster sample based on their spectral and $V-$band morphological classification. In order to be consistent with previous WINGS literature and the ongoing data release of OmegaWINGS, we also adopt the mass completeness limit of Log $(M_{*}/M_{\odot})=9.8$ \citep{pacc17}. Figure \ref{sfr-mstar} shows the SFR vs $M_{*}/M_{\odot}$ for the star-forming spirals in A3376 along with those from the control cluster sample. We use the SFR--$M_{*}$ best fit line for all the star-forming field galaxies in OmegaWINGS sample, and use it to bifurcate the \textit{purely} star-forming galaxies from the those in transition \citep[blue dashed line in Figure 2,][]{pacc16}. With our SFR--$M_{*}$ relation conforming to that of the whole OmegaWINGS data presented in \citet{pacc16}, most of the high-mass star-forming spirals are the blue spirals identified on the CMR, and having SFR--$M_{*}$ within 1.5$\sigma$ of the field best-fit line (blue solid line in Figure \ref{sfr-mstar}). This result has important connotations in furthering our current understanding of the nature of star formation in shock-heated environment of merging cluster systems. It demonstrates that at a low-$z$ minor cluster mergers with moderate shock intensity, such as A3376, could be capable of sustaining star formation in spiral galaxies. However, the magnitude of such ongoing star formation is not quite enough to be identified as purely star-forming galaxies (above the dashed line) thereby ruling out further quenching or  enhancement occurring due to the passage of the merger shock $~0.6$ Gyr ago. Utilising the data presented in Figure~\ref{sfr-mstar}, we perform a bootstrapping test by randomly extracting star-forming cluster spirals from OmegaWINGS control sample, to form samples with numbers equivalent to those of star-forming spirals in A3376. Repetitively populating the SFR--$M_{*}/M_{\odot}$ plane with these random test samples, we establish that there is a 95\% probability of the randomly drawn sample of star-forming cluster spirals from relaxed clusters at similar redshifts to result in the observed distribution of A3376 spiral galaxies in Figure~\ref{sfr-mstar}. The observed tightness in the SFR--$M_{*}/M_{\odot}$ relation of the star-forming spirals in A3376, therefore, is majorly due to lower number statistics. We however stress that our analysis included only the regular star-forming galaxies and not the emission-line star-forming spirals or starbursts (refer to Section~\ref{spec-dist}) which usually have higher star formation. Although it is beyond the scope of this paper, a dedicated study of these emission-line star-forming spirals is essential in order to establish a complete picture of plausible effect the merger shock is having on the overall star formation activity in cluster galaxies. Moreover, the spectroscopic observations of galaxies suffer from aperture effects as the fibre spectrograph targets only the central few kpc of cluster late-types at $z\sim0.046$. This therefore reveals  the star formation in the central region of galaxies although we cannot comment on the precise nature of star-forming activity in galaxy discs.  

Our analysis presents evidence that the star formation remains unchanged in the spiral galaxies of A3376 despite residing in shock influenced cluster environment. This is supported by the incidence of massive star-forming blue spirals located in projection to the central region of the cluster. Furthermore, we also observe the presence of at least two different populations of post-starburst galaxies, one of which is younger, blue and low mass. Collating the results we have so far, we thus propose that the merger shock which propagated at the onset of merger $\sim 0.6$ Gyr back may have a significant impact on the star formation in cluster galaxies, although different effects are observed for high- and low-mass transient galaxies like PSBs.          

\section{Discussion: Ongoing star formation in young post-merger cluster systems}
\label{discussion}

Amalgamating the whole picture, our results have established the existence of a dual environmental influence on the galaxies in A3376: the effect of immediate cluster environment since before the merger and the possible aftermath of the cluster merger occurring $\sim0.6$ Gyr ago. This is reinforced by the presence of passive spirals in regions of high density co-existing with a small population of blue star-forming spirals with relatively higher stellar masses lying in the regions of A3376 along which the merger shock passed through as early as $\sim0.6$ Gyr back. Furthermore, though their positions in the cluster cannot be ascertained, we also observe PSBs displaying at least two different formation ages, the youngest of which are low-mass blue PSBs with late-type morphologies.

Several studies have shown that the galaxy environment is more efficient in affecting the star formation in low mass galaxies \citep{ypeng10}, with transformations happening through preprocessing \citep{haines13} or through first infall \citep{jaffe15}. Our system, however, seems to display three broad spiral galaxy populations: a) passive galaxies already sitting on the red sequence spanning a broad range of masses, b) the usual `blue cloud' spirals having relatively low masses and c) a small but distinct group of massive blue star-forming spiral galaxies characteristic only to the central region of the merger system. These trends suggest a plausible scenario comprising two-fold environmental effects: a pre-merger environment which would result in spirals residing in the denser parts of their parent clusters getting redder due to truncation of star formation caused by ICM interactions, and a post-merger environment in which spirals are subjected to the heated ICM and expanding shock front since $\sim$0.6 Gyrs. The massive blue star-forming spirals thus may be the beacons of the latter. An estimate of the timescale and magnitude of such activity would then ascertain whether these trends really are the post-merger signatures on the star formation of spiral galaxies or whether they are peripheral spiral galaxies simply redistributed without any major observable changes. 

Simulations show that a typical $10^{10} M_\odot$ disc galaxy could have its star formation enhanced by more than a magnitude, and sustained for $\sim 100$ Myrs, as a result of a passing ICM shock (e.g. ram-pressure in dense cluster cores) with a pressure $\sim 10^{-22}$ N/m$^2$ \citep{kapferer09}. This advocates the possibility that the progressing shock compresses the gas in the galaxy causing it to form stars at a sustained level \citep[See also][]{sobral15}. Comparing this with the merger-shock pressure inferred from X-ray observations of A3376 within and across the radio relics \citep[$\sim 10^{-15}$ to $10^{-11}$  N/m$^{2}$;][]{akamatsu12}, we affirm that the impact of the shock could potentially ignite star formation in disc galaxies when subjected to a shock wave with $v_{s}\sim$1630 km/s \citep{urdampilleta18} resulting from the A3376 cluster merger. Our study thus contributes to and validates the general consensus that the time since the merger or the `age' and the magnitude of merger shock is crucial in determining the consequences on star formation in cluster galaxies.

Studies have also shown that galaxies run over by ICM shock are expected to display `Jellyfish'-like morphology, showing signatures of ongoing gradual strippping between $\sim 10$ Myrs to few $100$ Myrs after the passage of shock \citep{roediger14}. Interestingly, two of the aforementioned core blue star-forming spirals are also identified as `Jellyfish' candidates by \citet{poggianti16}. As a part of Gas And Stripping Phenomenon (GASP) survey \citep{gasp1}, IFU observations of one of the jellyfishes, JW108, show the gas being stripped with the galaxy showing a truncated disc \citep{gasp9}. The merger shock in A3376, though efficient in seemingly continuing the star formation in cluster spirals, however, is not strong enough at any point since $\sim0.6$ Gyr ago to initiate shock-pressure stripping of gas in cluster spirals. We thus corroborate the findings of \citet{gasp9} who argued that the orbits of such `post-stripped' galaxies somehow bring them in the densest regions where ram-pressure is extreme and in process get decelerated. We, therefore, propose an alternative interpretation that the stripping of gas in JW108 could also stand out as a case where a cluster spiral galaxy, which has already been experiencing dense cluster environment prior to $\sim0.6$ Gyr ago, is suddenly subjected to high-density ICM due to the onset of cluster merger thereby causing it to lose gas from the inner disc. The impact of the outgoing shock is thus realised only in the nuclear star formation in these jellyfishes and not in the gas-stripping signatures. This hypothesis is concurrent with the findings of \citet{mansheim17-a} who discover quenched galaxy population in the collision front of an ongoing merger, and \citet{ruggiero19} whose simulation of the Abell 901/2 system determined that jellyfish galaxies are preferentially found near the spatial boundaries arising from the confrontation between gas moving along the cluster and that from the remainder of the system, where the ram-pressure is thousand times intense \citep[See also][]{owers12,ebeling19}.

The incidence of PSBs in merging cluster systems have also been studied through simulations presented by \citet{bekki10}, who concluded that a 3:1 cluster-mass merger could lead to synchronised star-bursting activity in galaxies which later evolve to be PSBs. Furthermore, they argue that the PSBs show markedly different spatial distribution than the rest of the cluster population. Our study corroborates their findings with respect to a correlation observed between recent post-starburst activity and the time since the pericentric passage of the E and W clusters. However, the lack of significant number of PSBs limits us from infering anything robustly regarding their spatial distribution and number fractions; though the young PSBs do appear to lie within the region of influence along the merger axis. Our current understanding of PSBs in cluster environment presents them as a transition population going from star-forming to a passive state. However, recent census agree that the evolutionary pathways which lead to post-starburst stage in galaxies are varied in different environments \citep{pawlik19}. We interpret the existence of young low-mass PSBs in A3376 to be attributed to an episodic star formation either due to the shock pressure or the sudden surge in the ICM pressure at the instance of merger followed by quenching resulting from the ambient shock heated ICM. Thus, the PSB signatures instilled by the merging event are observable within a limited time period post merger, though a fraction of these PSBs are likely to be resulting from infall, independent of cluster-merger. The evidence that most of these young PSBs have late-type morphologies further supports a quenching mechanism acting on rapid timescales.

Such a two-fold environmental effect is perhaps studied in such detail for the first time in any post-merger system. Being a unique cluster with symmetric radio relics in itself and it's relatively low-$z$ puts A3376 in a very pivotal place in the current literature of known merger systems. With conflicting observations supporting quenching \citep{pranger14, mansheim17-b} or triggering of star formation \citep{stroe15a,stroe15b} in merging clusters with shocks, our study establishes distinct pre- and post-merger cluster environment at play. Moreover, our results imply that these contradictory observations might be the ramifications of the corresponding physical processes albeit observed at different times during the merging process, with a critical dependence on the magnitude of the aftermath of merging processes like e.g. strength of the merger shock. Despite most of the merger systems studied with respect to galaxy populations exhibit stronger shocks ($>$1630 km/s), are at high redshift and mostly major mergers, our investigation stresses the credibility of a minor merger such as A3376 system with moderate shocks to successfully reveal the pre-, ongoing and post-merger signatures on the star formation of member galaxies.     

\section{Conclusions}
\label{conclusions}
We present an in-depth analysis of star formation properties of galaxies in a nearby ($z\sim0.046$) young ($\sim$0.6 Gyr) post-merger cluster system A3376, with observed shock front. Exploiting the spectroscopic derivatives from the OmegaWINGS survey and the associated photometric information, our investigations were able to segregate, for the first time, the likely effects of the dynamic post-merger environment on the member galaxies from the pre-merger putative cluster environment. In this regard, our key results are : 
\begin{itemize}
 \item A3376 displays the vestiges of pre-merger relaxed cluster environment through the existence of passive spiral galaxies located in the central regions of the cluster between the two BCGs.
 \item We discover A3376 to contain a population of massive (Log$(M_{*}/M_{\odot})>10$) blue regular star-forming spirals in the cluster core. Moreover, we report an overall sustained star formation in cluster spirals similar to those in relaxed clusters at the same epoch, at fixed stellar masses. However with these galaxies located in the regions where the influence of merger shock was longest, we conclude that such a shock does not seem to affect the star formation in the cluster spirals of A3376.
 \item On the other hand, the observed low-mass (Log $(M_{*}/M_{\odot})\leq 10$) late-type blue PSBs could either be formed as a result of rapid quenching of low-mass spirals following the shock-induced star formation or due to the intense surge in the ICM pressures at the beginning of the merger.  
\end{itemize}

In a nutshell, spiral galaxies in merging cluster A3376 continue forming stars, at the very least, despite being under the influence of the outgoing shock front of moderate velocity ($\sim$1630 km/s). With the possibility of the shock front affecting high- and low-mass spirals differently, our results bridge the seemingly contradictory results observed in known merging cluster systems so far and establish that different environmental effects are at play right from pre- to post-merger stage. Nonetheless, our results necessitate the need for more systematic explorations of all the galaxy populations of such merger systems to truly constrain the effect of dynamic transient environments on the star formation properties of galaxies within.          

\section*{Acknowledgements}

REGM acknowledges support from the Brazilian agency \textit {Conselho
Nacional de Desenvolvimento Cient\'ifico e Tecnol\'ogico} (CNPq)
through grants 303426/2018-7 and 406908/2018-4. RMO  thanks the financial support provided by CAPES. A. M. acknowledges funding from the agreement ASI-INAF n.2017-14-H.0. J.F. acknowledges financial support from the UNAM- DGAPA-PAPIIT IN111620 grant, M\'{e}xico. B.V and M.G. acknowledge financial contribution from the grant PRIN MIUR 2017 n.20173ML3WW\_001 (PI: Cimatti); B.V. A.M, B.M.P, D.B. and M.G. acknowledge financial contribution from the INAF main-stream funding programme (PI: Vulcani). 


\bibliographystyle{mnras}
\bibliography{kk_paper4}






\bsp	
\label{lastpage}
\end{document}